\documentclass[useAMS,usenatbib,twocolumn]{mn2e}
\def\OutputDriver{dvips}

\usepackage[\OutputDriver]{graphicx}
\usepackage{times} 
\usepackage{amsmath}
\usepackage{rotating}
\usepackage{amssymb}
\usepackage{bm}

\usepackage{color}

\newcommand{\sm}{\sc}

\def\apjl{ApJ}
\def\mnras{MNRAS}

\def\physrep{Phys. Rep.}
\def\nat{Nature}

\def\prd{Phys. Rev. D}

\title[Streams and caustics]
      {Streams and caustics: the fine-grained structure of $\Lambda$CDM haloes}
      \author[M. Vogelsberger and S.D.M. White] {\parbox{17.5cm}{ Mark
          Vogelsberger$^{1,2}$\thanks{mvogelsberger@cfa.harvard.edu}, 
          Simon~D.~M.~White$^{1}$ 
        }\vspace{0.3cm}\\ $^1$Max-Planck-Institut f\"{u}r Astrophysik,
        Karl-Schwarzschild-Stra\ss{}e 1, 85740 Garching bei
        M\"{u}nchen, Germany\\
        $^{2}$Harvard-Smithsonian Center for Astrophysics, 60 Garden Street, Cambridge, MA 02138, USA}

\begin{document}

\pubyear{2010}

\pagerange{\pageref{firstpage}--\pageref{lastpage}} 

\maketitle

\label{firstpage}
\begin{abstract}
We present the first and so far the only simulations to follow the
fine-grained phase-space structure of galaxy haloes formed from generic
$\Lambda$CDM initial conditions. We integrate the geodesic deviation equation
in tandem with the N-body equations of motion, demonstrating that this can
produce numerically converged results for the properties of fine-grained
phase-space streams and their associated caustics, even in the inner regions
of haloes. Our effective resolution for such structures is many orders of
magnitude better than achieved by conventional techniques on even the largest
simulations. We apply these methods to the six Milky Way-mass haloes of the
Aquarius Project. At 8 kpc from halo centre a typical point intersects about
$10^{14}$ streams with a very broad range of individual densities; the $\sim
10^6$ most massive streams contribute about half of the local dark matter
density. As a result, the velocity distribution of dark matter particles
should be very smooth with the most massive fine-grained stream contributing
about 0.1\% of the total signal.  Dark matter particles at this radius have
typically passed 200 caustics since the Big Bang, with a 5 to 95\% range of 50
to 500. Such caustic counts are a measure of the total amount of dynamical
mixing and are very robustly determined by our technique. The peak densities
on present-day caustics in the inner halo almost all lie well below the mean
local dark matter density.  As a result caustics provide a negligible boost
($<0.1\%$) to the predicted local dark matter annihilation rate. The effective
boost is larger in the outer halo but never exceeds about $10\%$. Thus
fine-grained streams and their associated caustics have no effect on the
detectability of dark matter, either directly in Earth-bound laboratories, or
indirectly through annihilation radiation, with the exception that resonant
cavity experiments searching for axions may see the most massive local
fine-grained streams because of their extreme localisation in
energy/momentum space.
\end{abstract}


\section{Introduction}

Dark matter is supposed to be the principal driver of structure formation in
the Universe, and a whole industry has developed over the last few decades
searching for the still elusive dark matter particle.  Particle physics has
provided some well-motivated candidates. Among them, the neutralino, a weakly
interacting massive particle (WIMP) associated with a supersymmetric extension
of the standard model of particle physics, is currently favoured
\citep[see][for
  reviews]{1996PhR...267..195J,2005PhR...405..279B,2009NJPh...11j5006B}.
Although this particle interacts with standard model particles only weakly, it
may nevertheless be detectable.  Recently the dark matter community has been
stimulated by a variety of observed ``anomalies'' in cosmic-ray signals. Among
these are: (i) an excess in the positron fraction recently (re)measured by the
PAMELA experiment \citep[][]{2009Natur.458..607A}; (ii) an excess in the total
flux of electrons and positrons measured by the FERMI satellite
\citep[][]{2009PhRvL.102r1101A}; (iii) an excess in diffuse microwave
radiation in the general direction of the Galactic Centre 
\citep[the ``WMAP haze'',][]{2007PhRvD..76h3012H}. Although these results may well find their
explanation in ordinary astrophysical processes (see, for example,
\cite{2009PhRvD..80f3005M} for a pulsar explanation of the PAMELA results), it
is alluring to relate them all to dark matter annihilation in the Galactic
halo.

In addition to these indirect signals, the DAMA/LIBRA dark matter detection
experiment has, for about a decade \citep[][]{2008EPJC..tmp..167B,
  2010EPJC...67...39B} seen a significant amplitude modulation in its detector
count rate. Although apparently incompatible with results from other
experiments \citep[see e.g.][]{2004PhRvD..70l3513S, 2005PhRvD..71l3520G,
  2006JPhCS..39..166G, 2009PhRvD..80k5008F}, the DAMA/LIBRA collaboration
interprets this as an annual modulation of the flux of dark matter particles
through their detectors.

The small-scale structure of Cold Dark Matter (CDM) haloes can, in principle,
substantially influence the signal both in indirect and in direct dark matter
detection experiments. Annihilation rates and detector count rates depend
strongly on the density and energy distributions of particles local to other
particles and to detector nuclei, respectively, and so are sensitive to
small-scale fluctuations in these distributions.  For example, for
cross-sections consistent with relic abundance constraints, the PAMELA data
can be explained through annihilation only if rates are 100 to 1000 times
those predicted for a locally smooth dark matter distribution, thus requiring
substantial ``boost factors'' \citep[e.g.][]{2008PhRvD..78j3520B}. Such boosts
could come from non-standard particle physics, or they might reflect
substantial inhomogeneities in the dark matter distribution on small scale. A
population of abundant, self-bound subhaloes of very low mass but very high
internal density has often been invoked in this context
\citep[][]{1999ApJ...524L..19M, 2005JCAP...08..003G,
  2005Natur.433..389D,2008Natur.454..735D}, but recent high-resolution
simulations suggest that such subhaloes are neither dense enough nor abundant
enough in the inner regions of $\Lambda$CDM haloes to have more than a minor
effect on the observable signatures of annihilation
\citep{2008Natur.456...73S}.

Another possible boost mechanism is related to the fine-grained structure of
dark matter haloes. Before the onset of nonlinear structure formation, dark
matter was almost uniformly distributed, with weak density and velocity
perturbations and with very small thermal motions. The particles thus occupied
a thin, space-filling and almost three-dimensional sheet in the full
six-dimensional phase-space. Subsequent collisionless evolution under gravity
stretched and folded this sheet, but did not tear it. Thus, the dark matter
distribution at a typical point in a present-day halo is predicted to be a
superposition of many fine-grained streams, each of which has a very small
velocity dispersion, has a density and mean velocity which vary smoothly with
position, and corresponds to material from the vicinity of a different point
in the linear initial conditions. Folds in the fine-grained phase-sheet give
rise to projective catastrophes known as caustics where the spatial density
and hence the annihilation rate is locally very high, limited only by the
small but nonzero thickness of the phase-sheet
\citep[e.g.][]{2001PhRvD..64f3515H,2008PhRvD..77d3531N}.  Although the
extraordinary improvement of N-body simulations in recent years has allowed
many aspects of the dark matter distribution at the solar position to be
predicted in considerable detail \citep[see, for
  example,][]{2008MNRAS.391.1685S,2008Natur.454..735D, 2009MNRAS.398L..21S},
such simulations are still very far from resolving the fine-grained
phase-space structure which gives rise to caustics. They are thus unable to
provide a realistic estimate of how much caustics boost annihilation in the
inner Galactic halo.

Fine-grained structure might also play a crucial role in the interpretation
and modelling of dark matter signals in laboratory detectors like
DAMA/LIBRA. It is unclear, for example, whether the Maxwellian usually assumed
describes the dark matter velocity distribution at the solar position
accurately. Indeed, recent simulation work has shown that a multivariate
Gaussian is likely a better description, and that the assembly history of the
Milky Way may be reflected in broad features in the particle energy
distribution \citep{2009MNRAS.395..797V}.  One may also wonder whether
individual fine-grained streams might eventually be visible in detector
signals as spikes at specific velocities. Axion detectors like ADMX, in
particular, have very high energy resolution, and could in principle
disentangle hundreds of thousands of streams. Again this is far beyond the
resolution limit of current analysis techniques applied to even the highest
resolution N-body simulations of halo formation.

In recent work we have developed an entirely new approach capable of following
the evolution of fine-grained streams and their associated caustics in fully
general simulations of halo formation. This approach is based on integrating
the geodesic deviation equation (GDE) in tandem with the N-body equations of
motion. For every simulation particle and at every timestep it returns the
spatial density and the velocity dispersion tensor of the fine-grained stream
in which the particle is embedded. This in turn allows all passages of the
particle through a caustic to be identified and recorded. In
\cite{2008MNRAS.385..236V} we presented this GDE scheme and tested it on
orbits in fixed potentials and on N-body simulations of static equilibrium
haloes.  In \cite{2009MNRAS.400.2174V} we then applied the scheme to a
simplified model of CDM halo formation: collapse from spherically symmetric,
self-similar and linear initial conditions. This showed that earlier work
based on spherically symmetric similarity solutions
\cite[e.g.][]{2006PhRvD..73b3510N,2007JCAP...05..015M} was unrealistic because
these solutions are violently unstable to nonradial perturbations. These
substantially alter the stream and caustic structure. Finally in
\cite{2009MNRAS.392..281W} we showed how the GDE scheme allows the
annihilation enhancement due to caustics and other fine-grained structure to
be calculated explicitly by time-integration along particle orbits rather than
by spatial integration over the particle distribution. The current paper is
the culmination of this programme and analyses for the first time the
fine-grained structure of haloes forming from fully general $\Lambda$CDM
initial conditions.  We resimulate haloes from the Aquarius Project, which
studied six Milky Way mass objects at various numerical resolutions. The
coarse-grained structure of these haloes has already been studied in
considerable detail in previous papers, e.g. the subhalo population in
\cite{2008MNRAS.391.1685S} and \cite{2008Natur.456...73S}, the dark matter
distribution in the inner regions in \cite{2009MNRAS.395..797V} and various
radial profiles in \cite{2010MNRAS.402...21N}.

The plan of our paper is as follows. In Section 2 we briefly describe the
initial conditions and the numerical approach we use to resolve the
fine-grained phase-space structure. Section 3 begins by presenting results on
one of the Aquarius haloes at a variety of resolutions. We demonstrate that,
for fixed gravitational softening, convergent results can be
  obtained, free of significant discreteness or two-body relaxation
  effects. We then analyse the implications for dark matter detection of the
structure predicted for fine-grained streams and caustics. Finally, we use our
full halo sample to tackle the question of how much scatter in fine-grained
properties is expected. We make some concluding remarks in Section
4. An Appendix shows that although discreteness effects are
  negligible in our high resolution simulations, their predictions for
  fine-grained stream densities (but not for caustic structure) are influenced
  by gravitational softening because this affects the tidal forces on
  particles passing close to halo centre.

\section{Initial conditions and numerical methods}

We resimulate the six Milky Way-mass haloes of the Aquarius Project
\citep[][]{2008MNRAS.391.1685S} using our GDE technique to follow the
fine-grained phase-space evolution in detail.  The cosmological parameters
assumed for these $\Lambda$CDM simulations are
$\Omega_{m0}=0.25,\Omega_{\Lambda0}=0.75,\sigma_8=0.9,n_s=1$ and $H_0=73~{\rm
  km}~{\rm s}^{\rm -1}~{\rm Mpc}^{\rm -1}$, where all quantities have their
standard definitions. The haloes were selected at $z=0$ based on their mass,
and were required to have no close massive companion at that time; they are
named Aq-A to Aq-F. Each halo was resimulated at a variety of numerical
resolutions, indicated by a number $n$ in its full name, e.g. Aq-A-3. The
resolution levels differ in the particle mass and softening length employed,
with 1 designating the highest resolution and 5 the lowest. In the following
we will use the same naming convention and mass resolution as in the original
papers but different softening lengths. This is because accurate integration
of the geodesic deviation equation (GDE) requires a larger softening length to
achieve stable results than does integration of the particle trajectories
themselves. Unless otherwise stated, we use a constant comoving
Plummer-equivalent softening of $3.4$~kpc in all our
simulations. The Appendix examines the effect of softening on
  our results in some detail. All other simulation parameters are the same as
  in \cite{2008MNRAS.391.1685S}, to which we refer the reader for further
  technical information.

Our experiments are carried out with the {\sm P-GADGET-3} code
\citep[][]{2005MNRAS.364.1105S} with the GDE modifications described in
\cite{2008MNRAS.385..236V} and \cite{2009MNRAS.400.2174V}, where we focused on
a description of the relevant equations in physical coordinates and applied
them to static potentials, to isolated equilibrium haloes and to halo
formation from self-similar, spherically symmetric initial conditions. The
simulations of this paper take into account the full $\Lambda$CDM framework,
using the cosmological parameters given above, and are carried out in comoving
coordinates. We therefore begin by describing how the GDE and the associated
tensors can be transformed from physical to comoving coordinates, and how this
is implemented in our simulation code.  For completeness, we also describe how
the physical stream density is calculated and how caustic passages can be
identified.

The time-dependent transformation to the comoving frame used by our simulation
code is given by\footnote{We use the following notation to
clearly distinguish between three- and six-dimensional quantities: an
underline denotes a $\mathbb{R}^3$ vector and two of them denote a
$\mathbb{R}^{3 \times 3}$ matrix. An overline denotes a $\mathbb{R}^6$ vector and two of
them denote a $\mathbb{R}^{6 \times 6}$ matrix. See also \cite{2008MNRAS.385..236V} and \cite{2009MNRAS.400.2174V}.}
\begin{equation}
\underline{x}(\underline{x}^\prime, \underline{v}^\prime) = a \underline{x}^\prime, \quad
\underline{v}(\underline{x}^\prime, \underline{v}^\prime) = H a \underline{x}^\prime + \frac{1}{a} \underline{v}^\prime,
\end{equation}
where $H=\dot{a}/a$ denotes the Hubble parameter, $a$ the scale factor, and
comoving coordinates are primed. In physical coordinates the distortion tensor
is defined as
\begin{equation}
 \overline{\overline{D}} = \frac{\partial \overline{x}}{\partial \overline{q}} =
 \left(
   \begin{array}{cc} 
     \partial \underline{x} / \partial \underline{q}   &  \partial \underline{x} / \partial \underline{p}\\
     \partial \underline{v} / \partial \underline{q}   &  \partial \underline{v} / \partial \underline{p}\\    
   \end{array}
\right)=
 \left(
   \begin{array}{cc}
     \underline{\underline{D}}_{xq}   &  \underline{\underline{D}}_{xp}\\
     \underline{\underline{D}}_{vq}   &  \underline{\underline{D}}_{vp}\\
   \end{array}
\right),
\end{equation}
where $\underline{q}$ and $\underline{p}$ denote the initial position and
velocity (in physical coordinates), respectively.  It describes how a local
phase-space element in physical coordinates is deformed along the trajectory
of the particle while conserving its volume in order to obey Liouville's
theorem. In comoving coordinates the distortion tensor is accordingly defined
as
\begin{equation}
 \overline{\overline{D}}^\prime \!\!\!=\!\! \frac{\partial \overline{x}^\prime}{\partial \overline{q}^\prime} \!\! =\!
 \left(
   \begin{array}{cc}
     \partial \underline{x}^\prime / \partial \underline{q}^\prime   &  \partial \underline{x}^\prime / \partial \underline{p}^\prime\\
     \partial \underline{v}^\prime / \partial \underline{q}^\prime   &  \partial \underline{v}^\prime / \partial \underline{p}^\prime\\ 
   \end{array}
\right)\!\! =\!\! 
 \left(
   \begin{array}{cc}
     \underline{\underline{D}}_{x^\prime q^\prime}^\prime   &  \underline{\underline{D}}_{x^\prime p^\prime}^\prime\!\!\\
     \underline{\underline{D}}_{v^\prime q^\prime}^\prime   &  \underline{\underline{D}}_{v^\prime p^\prime}^\prime\!\!\\
   \end{array}
\right),
\end{equation}
where all phase-space coordinates are now expressed in comoving coordinates.
The relation between physical and comoving distortion tensors can be derived
from the differential relations, yielding
\begin{equation}
\overline{\overline{D}}(t) =
\overline{\overline{C}}_{\overline{x}^\prime \rightarrow \overline{x}}(t)
\overline{\overline{D}}^\prime(t)
\overline{\overline{C}}_{\overline{x} \rightarrow \overline{x}^\prime}(t_{\rm initial}),
\end{equation}
where we have defined the two transformation tensors
\begin{equation}
\overline{\overline{C}}_{\overline{x}^\prime \rightarrow \overline{x}} 
=
\frac{\partial \overline{x}}{\partial \overline{x}^\prime} 
=
\left(
\begin{array}{cc}
a \underline{\underline{1}} & \underline{\underline{0}}\\
H a \underline{\underline{1}} & a^{-1} \underline{\underline{1}}
\end{array}
\right),
\end{equation}
and
\begin{equation}
\overline{\overline{C}}_{\overline{x} \rightarrow \overline{x}^\prime} =
\frac{\partial \overline{x}^\prime}{\partial \overline{x}}  
=
\left(
\begin{array}{cc}
a^{-1} \underline{\underline{1}} & \underline{\underline{0}}\\
-H a\underline{\underline{1}} & a \underline{\underline{1}}
\end{array}
\right).
\end{equation}
We note that one of these transformation tensors is evaluated at $t$ whereas
the other is evaluated at the initial time $t_{\rm initial}$.  The reason for
this is the time-dependence of the coordinate transformation in Eq. 1,
where both the scale factor and the Hubble parameter change with
time. Liouville's theorem guarantees that
$\mathrm{det}(\overline{\overline{D}}(t))=\mathrm{det}(\overline{\overline{D}}^\prime(t))=1$, 
so the comoving distortion tensor has the same conserved
determinant as the physical one. This means that local phase-space elements in
the comoving frame conserve both volume and orientation as the system
evolves.

\begin{figure}
\center{
\includegraphics[width=0.45\textwidth]{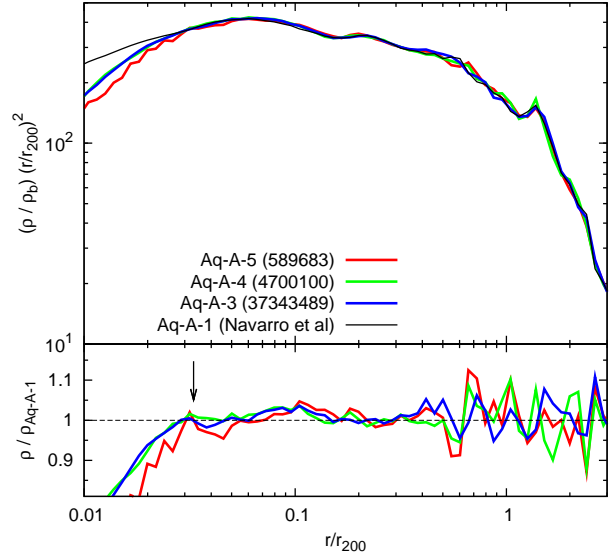}
}
\caption{Spherically averaged density profiles of Aq-A-3,4 and 5 at $z=0$. The
  comoving softening length of all simulations is the same ($\epsilon=3.4~{\rm
    kpc}$; note that $r_{200}=246$~kpc for this halo) but the particle number
  increases from Aq-A-5 to Aq-A-3 as shown by the values of $N_{200}$ in the
  figure. The lower panel shows the difference between these profiles and that
  presented by \protect\cite{2010MNRAS.402...21N} for Aq-A-1, 
    for which $N_{200}\sim 10^9$.  The convergence between the different
  resolutions is very good. All begin to fall below Aq-A-1 for $r < 0.025
  r_{200} = 6$~kpc because of the larger softening of our GDE
  resimulations. An arrow in the lower panel indicates 8~kpc, the radial
  position of the Sun within the Milky Way.}
\label{fig:mean_density} 
\end{figure}

To integrate the comoving distortion tensor, we need to derive its equations
of motion. We start with the physical GDE equation
\begin{equation}
\dot{\overline{\overline{D}}}=\overline{\overline{T}}~\overline{\overline{D}},
\end{equation}
and introduce the peculiar potential field
\begin{equation}
\phi = a \Phi + \frac{a^2 \ddot{a}}{2} \underline{x}^{\prime 2},
\end{equation}
which is related to the density field via Poisson's equation
\begin{equation}
\underline{\nabla}_{x^\prime}^{2} \phi = 4 \pi G \left(\rho^\prime(\underline{x}^\prime) - \rho_b^\prime\right),
\end{equation}
where $ \rho_b^\prime$ denotes the comoving mean background density and the
Laplacian is taken in comoving space. We note that the force field driving the
motion of simulation particles is also derived from this peculiar potential.
Since the GDE is directly related to the equations of motion of the particles
themselves, it is natural to introduce a peculiar configuration-space tidal
tensor $\underline{\underline{T}}^\prime$ with components $T_{ij}^\prime =
-\partial^2 \phi / \partial x_i^\prime \partial x_j^\prime$
\begin{equation}
\frac{1}{a^3} \underline{\underline{T}}^\prime = \underline{\underline{T}} - \frac{\ddot{a}}{a}\underline{\underline{1}}.
\end{equation}
We can use this tidal tensor to write the equations of motion for the
comoving distortion tensor in the following form
\begin{align}
\dot{\underline{\underline{D}}}_{x^\prime q^\prime}^\prime=\frac{1}{a^2} \underline{\underline{D}}_{v^\prime q^\prime}^\prime,
&\quad
\dot{\underline{\underline{D}}}_{x^\prime p^\prime}^\prime=\frac{1}{a^2} \underline{\underline{D}}_{v^\prime p^\prime}^\prime, \\ \nonumber 
\dot{\underline{\underline{D}}}_{v^\prime q^\prime}^\prime=\frac{1}{a} \underline{\underline{T}}^\prime\underline{\underline{D}}_{x^\prime q^\prime}^\prime,
&\quad
\dot{\underline{\underline{D}}}_{v^\prime p^\prime}^\prime= \frac{1}{a} \underline{\underline{T}}^\prime\underline{\underline{D}}_{x^\prime p^\prime}^\prime.  \nonumber 
\end{align}
Thus, the equations of motion for the comoving distortion tensor have
exactly the same form as those for the physical one. The only difference is
the appearance of the scale factor $a$ in the comoving equations.  The initial
conditions for these equations follow from those for the physical distortion
tensor, taking into account $\overline{\overline{C}}_{\overline{x}^\prime
  \rightarrow \overline{x}}~\overline{\overline{C}}_{\overline{x} \rightarrow
  \overline{x}^\prime}=\overline{\overline{1}}$
\begin{align}
\underline{\underline{D}}_{x^\prime q^\prime}^\prime (t_{\rm initial}) = \underline{\underline{1}}, 
&\quad
\underline{\underline{D}}_{x^\prime p^\prime}^\prime (t_{\rm initial}) = \underline{\underline{0}},  \\ \nonumber
\dot{\underline{\underline{D}}}_{x^\prime q^\prime}^\prime (t_{\rm initial}) = \underline{\underline{0}}, 
&\quad
\dot{\underline{\underline{D}}}_{x^\prime p^\prime}^\prime (t_{\rm initial}) = \underline{\underline{1}}.   \nonumber
\end{align}
Based on these equations we can construct kick- and drift-operators for the
leapfrog time integrator, similar to those for the position and velocity in
the comoving frame.

\begin{figure*}
\center{
\includegraphics[width=0.9\textwidth]{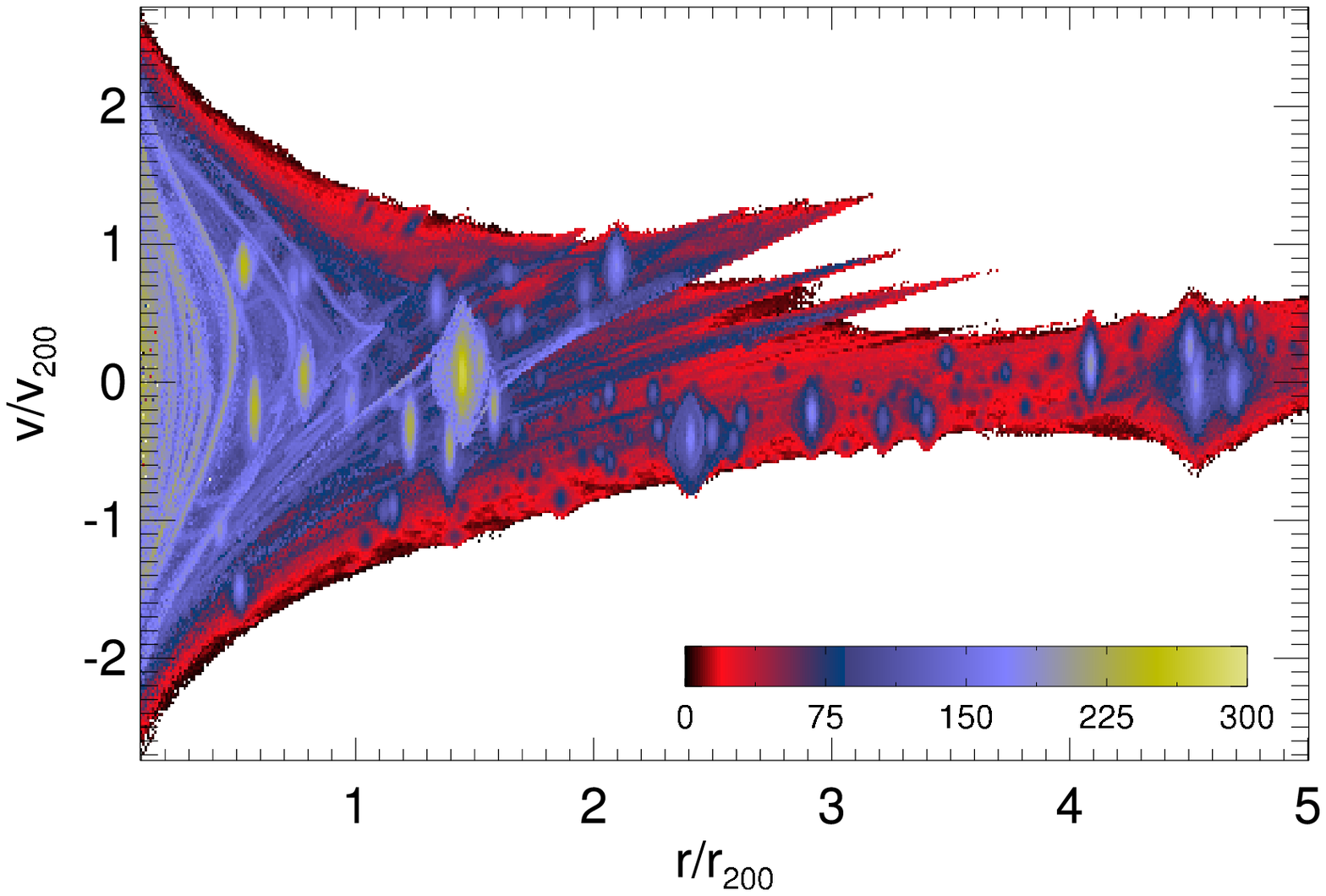}
\includegraphics[width=0.9\textwidth]{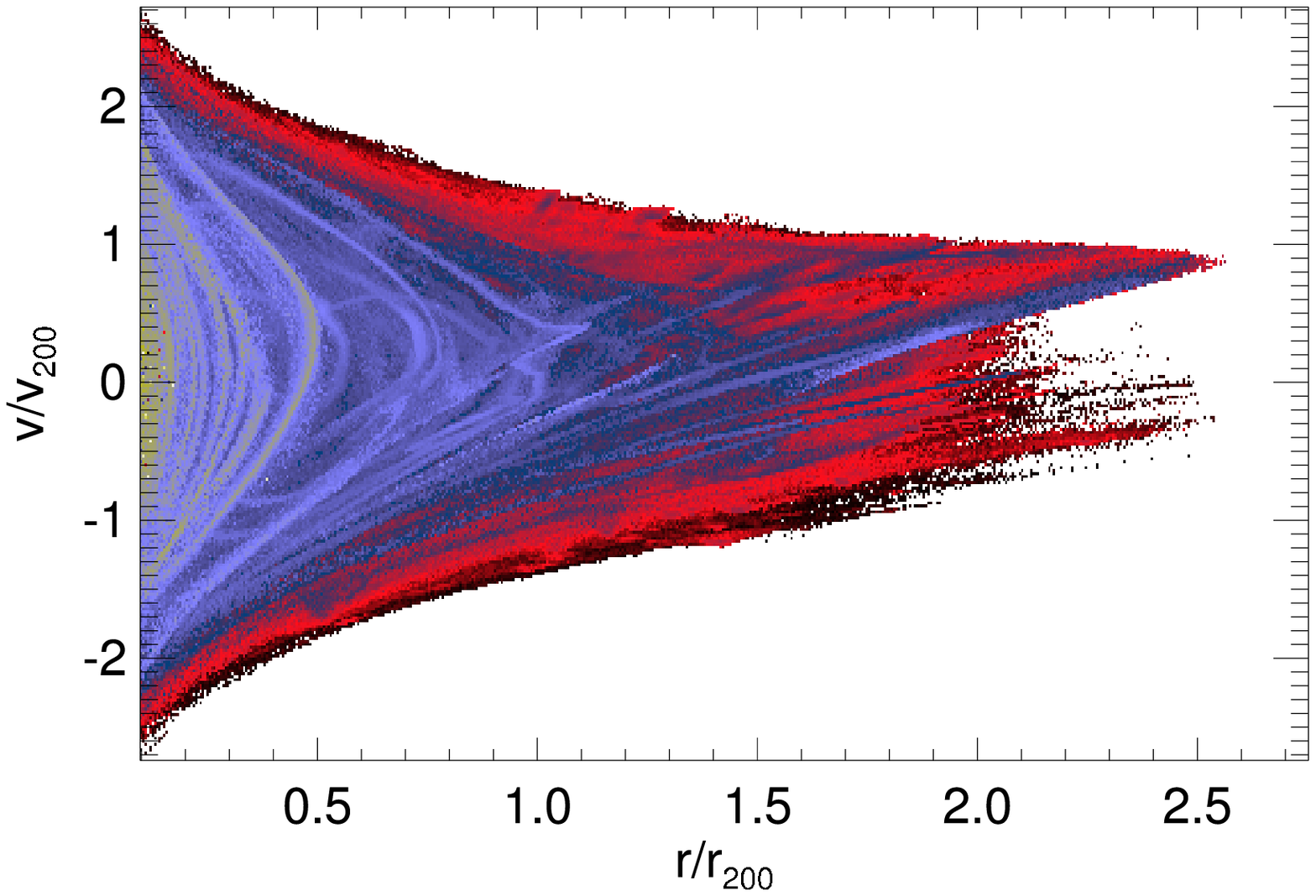}
}
\caption{Top panel: Phase-space structure of the Aq-A-3 halo at $z=0$.  Colour
  encodes the number of caustics passed by each individual simulation particle
  with the largest numbers plotted last. Subhaloes clearly stand out in this
  plot, because shorter orbital times lead to a higher number of caustic
  passages in their centres. We note that the streams visible in this plot are
  not fine-grained streams, but rather tidal streams resulting from disrupted
  subhaloes. Particles previously associated with subhaloes have higher
  caustic counts than other neighboring main halo particles, so they stand out
  in this phase-space portrait. Bottom panel: Same as the top panel, but with
  all bound subhaloes removed so that only the main halo component
  remains. Tidal streams from disrupted subhaloes are now more clearly
  visible. Note that this plot only shows particles bound to the main halo,
  while the top panel includes all particles within $5~r_{\rm 200}$. This is
  why the main halo contribution ends at about $2.5~r_{\rm 200}$. The colour
  scales for the two panels are the same, as indicated in the top panel.}
\label{fig:phasespace_caustics} 
\end{figure*}
\begin{figure*}
\center{
\includegraphics[width=0.9\textwidth]{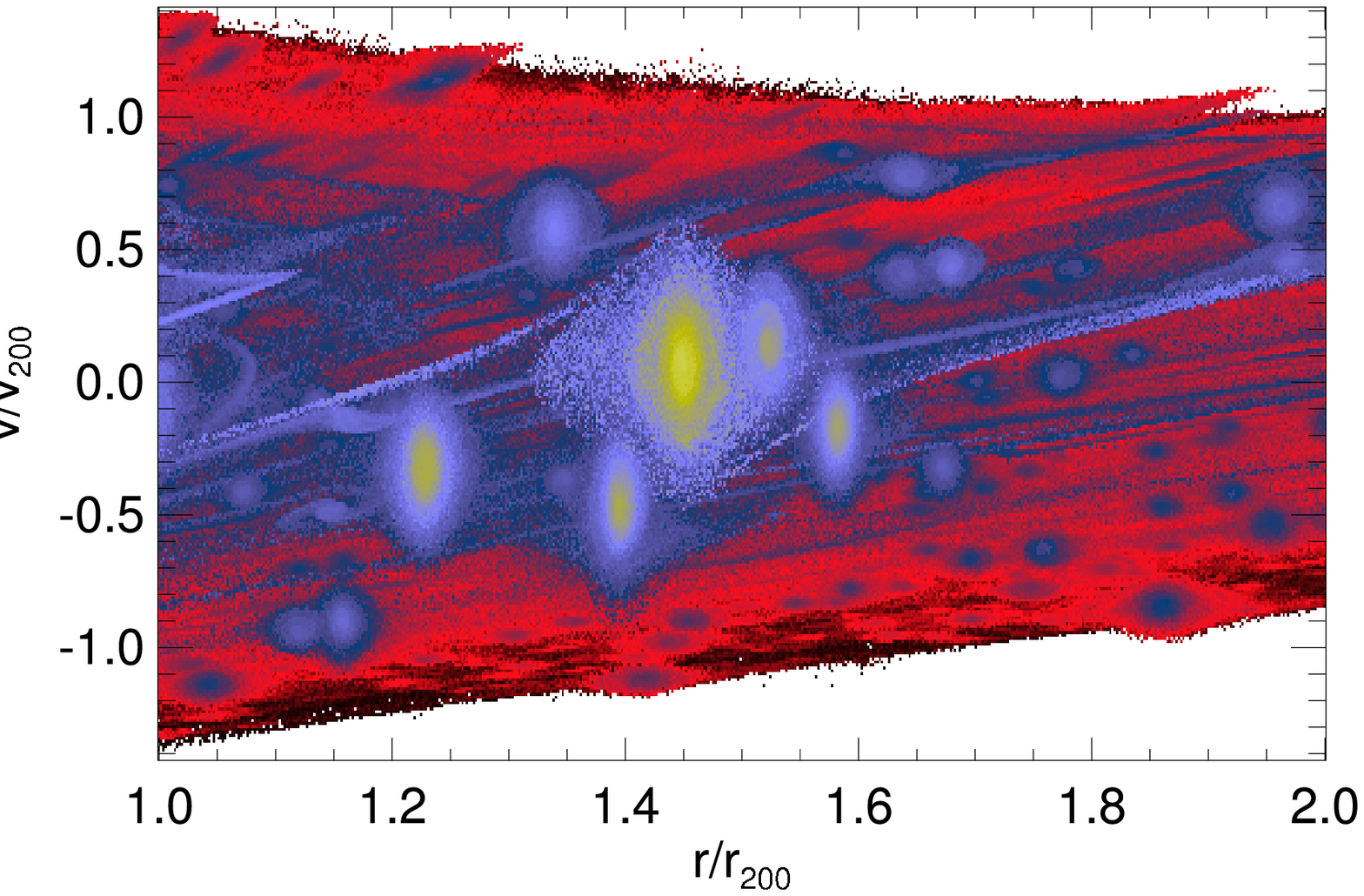}
\includegraphics[width=0.9\textwidth]{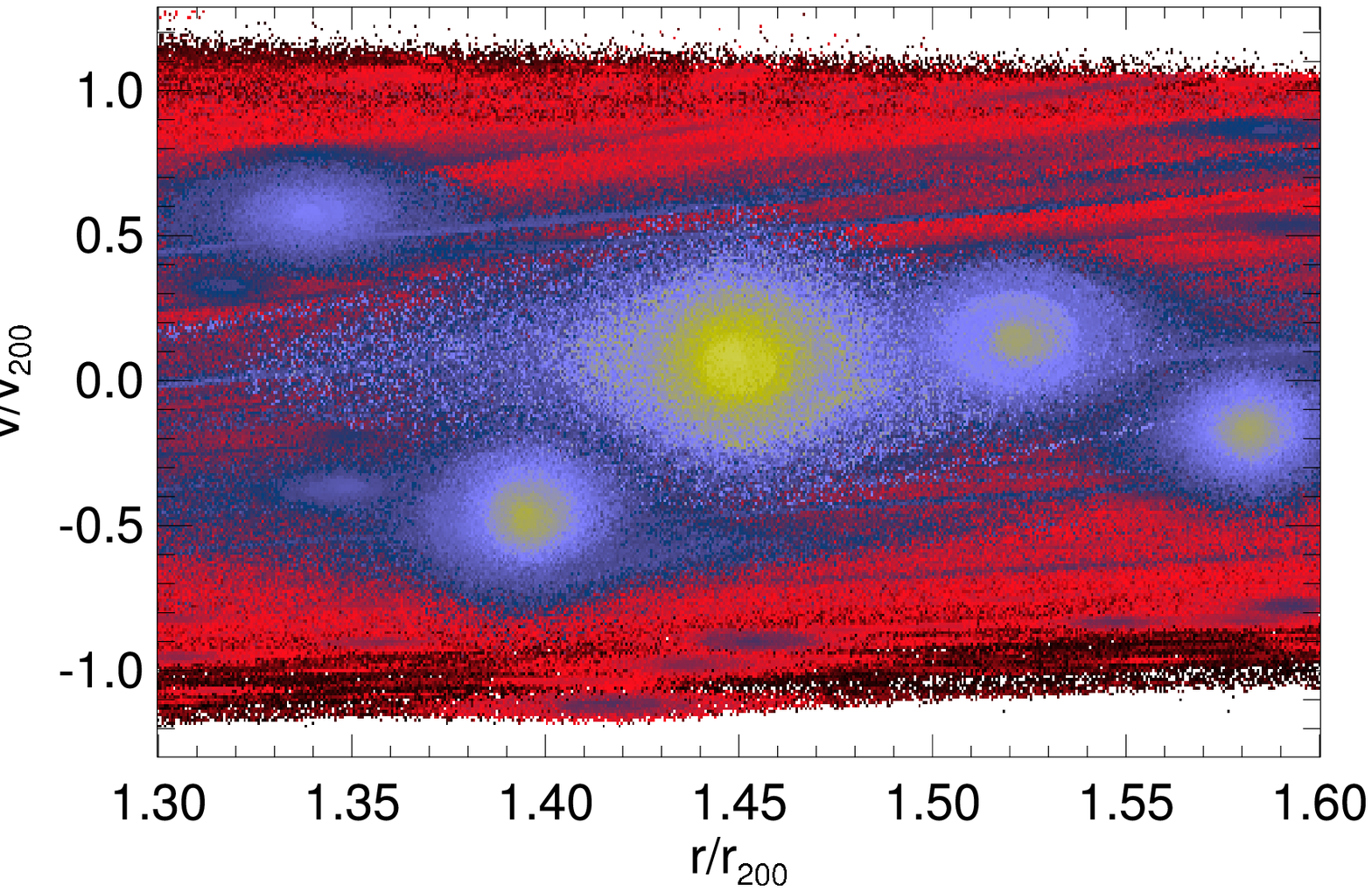}
}
\caption{Two magnifications of the phase-space structure of Aq-A-3 focusing on
  one of the prominent subhaloes. The colour scale is the same as in
  Fig.~\ref{fig:phasespace_caustics}. The tidal streams again stand out very
  clearly.}
\label{fig:phasespace_caustics_zoom} 
\end{figure*}

Let us now discuss how to derive stream densities from these comoving
quantities.  Each stream density is associated with a particular simulation
particle, and describes the local density of the fine-grained stream it is
embedded in.  Our goal is to derive an equation that yields the physical
stream density from the comoving distortion tensor discussed above.  To begin,
we note that the configuration-space distortion tensor can be derived from the
phase-space distortion tensor by applying two projection operators as follows
\begin{equation}
\underline{\underline{D}}=
\left(\underline{\underline{1}} \quad  \underline{\underline{0}} \right) 
\overline{\overline{D}}
\left(
   \begin{array}{c}
     \underline{\underline{1}} \\
     \underline{\underline{V}}_q
   \end{array}
\right),
\end{equation}
where $\underline{\underline{V}}_q= \partial{\underline V}/\partial{\underline
  q}$ is the spatial gradient of ${\underline V}({\underline q})$, the mean
initial sheet velocity as a function of initial position 
\citep[see][for a more formal definition]{2009MNRAS.392..281W}.
Caustic passages can be identified by sign changes in the determinant of this
tensor. We note that this is true both in physical and in comoving space.  The
physical stream density can be calculated from the volume stretch factor
implied by this linear transformation in physical configuration-space
\begin{equation}
\rho_s = \frac{\rho_{s,0}}{\left|
  \mathrm{det}\left(\underline{\underline{D}}\right)\right|}, 
\end{equation}
as we described in \cite{2008MNRAS.385..236V}. Using $\rho_{s,0} =
\rho_{s,0}^\prime/a(t_{\rm initial})^3$ and defining the peculiar velocity
gradient $\underline{\underline{V}}_{q^\prime}^\prime =
\underline{\underline{V}}_q - H(t_{\rm initial}) \underline{\underline{1}}$ we
can write the physical stream density as:
\begin{equation}
\rho_s = 
\frac{\rho_{s,0}^\prime}
{a^3\left| \mathrm{det}\left[  
\underline{\underline{D}}_{x^\prime q^\prime}^\prime +
  a(t_{\rm initial})^2 ~\underline{\underline{V}}_{q^\prime}^\prime ~
 \underline{\underline{D}}_{x^\prime p^\prime}^\prime\right]\right|}.
\end{equation}
At early times when the dark matter is nearly uniform, all velocities and
thus all velocity gradients are small and the second term in the determinant
in this equation is small compared to the first. Thus we may
approximate
\begin{equation}
\rho_s = 
\frac{\rho_{s,0}^\prime}
{a^3 \left| \mathrm{det}\left(  
\underline{\underline{D}}_{x^\prime q^\prime}^\prime \right)\right|}.
\end{equation}
\begin{figure*}
\center{
\includegraphics[width=0.825\textwidth]{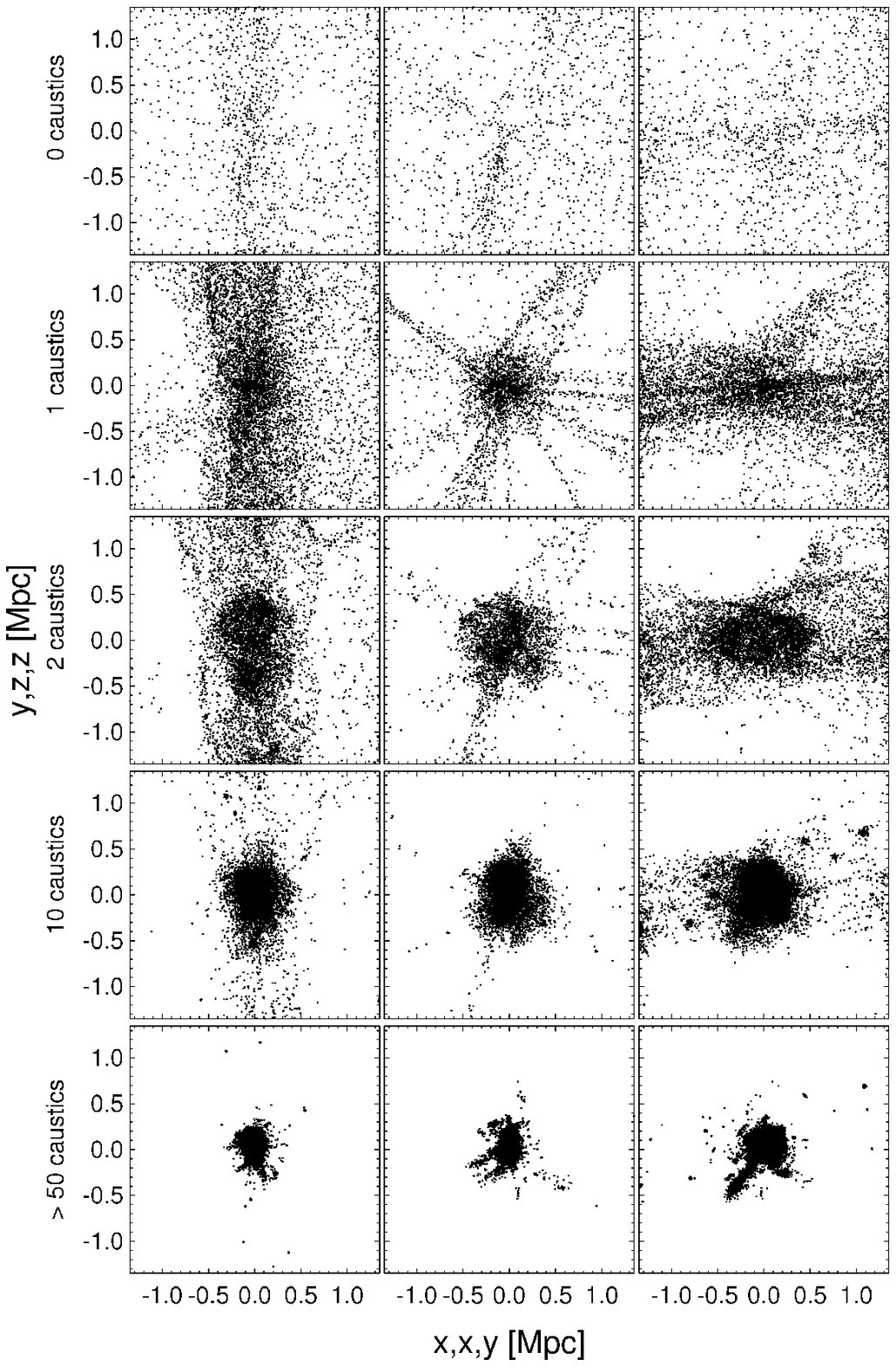}
}
\caption{Slices of thickness $0.1~{\rm Mpc}$ through (xy, xz, yz) projections
  of the particle distribution in Aq-A-5, filtered by the number of caustics
  passed.  The first four rows match the exact caustic passage number
  indicated, whereas the last row filters for particles that have passed more
  than $50$ caustics. The degree of structure in the different panels
  increases with the number of caustics passed.  Particles that have not
  passed any caustic form a smooth distribution compared to the other
  panels of the figure. }
\label{fig:caustic_nums} 
\end{figure*}
This equation is sufficiently accurate for our purposes and we will use
it below to calculate the stream densities associated with each particle
in our simulations. For consistency we will also neglect the (small)
initial variations in density and will assume the cosmological mean
value everywhere, i.e.
\begin{equation}
\rho_{s,0} = \frac{3 { H(t_{\rm initial})}^2}{8 \pi G}.
\end{equation}

\section{Quantifying fine-grained halo structure}

The results in the next few subsections are based on resimulations of the Aq-A
halo at resolution levels 3, 4 and 5.  We note again that our simulations use
a significantly larger softening length than the original Aquarius simulations
in order to mitigate the numerical noise sensitivity of the GDE.  Unless
otherwise stated, we use a Plummer-equivalent comoving softening length of
$3.4~{\rm kpc}$.  Note that we do not change softening between the different
resolution levels that we are going to discuss.  We will demonstrate that we
reach convergence at level 4 with this set-up. Additional experiments show
that this is the smallest softening length for which we could achieve
convergence at this resolution level; smaller values require a larger particle
number to converge. In the Appendix we show that our results
  for fine-grained stream densities (but not for caustic counts) are sensitive
  to the assumed softening because of its effect on the central structure of
  halo and subhalo potentials. Where needed, we also assume a dark matter
particle with a velocity dispersion of $0.03~{\rm cm}/{\rm s}$ today.  This
corresponds to a neutralino of mass $100~{\rm GeV/c^2}$ that decoupled
kinetically at a temperature around $10$~MeV.

To get started we show in Fig.~\ref{fig:mean_density} the spherically averaged
density profile of the Aq-A halo at redshift $z=0$ for all three resolution
levels. Note that $r_{200}= 246$~kpc for this halo.  Convergence is excellent
over the full radial range plotted, as is more obvious in the lower panel
where we plot the difference between our profiles and that given by
\cite{2010MNRAS.402...21N} for the highest resolution simulation Aq-A-1.
Softening clearly effects all three of our simulations similarly and only in
the innermost regions; the deviation is at the percent level at the radius
corresponding to the Sun's position within the Milky Way, but
reaches 10\% by 4 kpc. \cite{2010MNRAS.402...21N} made similar convergence
tests for the radial profiles of a variety of quantities in the original
simulations.

\subsection{Caustic counts}

Our focus in this paper is not on the coarse-grained but on the fine-grained
structure of our haloes. We begin by looking at the distribution of the number
of caustics passed by particles in Aq-A.  Caustics are identified through
changes in the sign of the determinant of the comoving configuration-space
distortion tensor and are counted along the trajectory of each particle.  In
Fig.~\ref{fig:phasespace_caustics} (top panel) we show the distribution of
this caustic count in a phase-space diagram at $z=0$ for Aq-A-3. Colour
encodes the number of caustics passed by each particle, as indicated by the
colour bar. The particles were sorted by their caustic count before plotting,
so that the particle with the highest caustic count is plotted in each
occupied pixel. This phase-space diagram can be compared directly to the one
in \cite{2009MNRAS.400.2174V} which shows an isolated halo which grew from a
spherically symmetric and radially self-similar perturbation of an Einstein-De
Sitter universe.  In both cases the number of caustic passages increases
towards the centre of the halo, due to the shorter dynamical timescales in the
inner regions: caustic count is roughly proportional to the number of orbits
executed by each particle, because caustics occur near orbital turning
points. The most important difference between the isolated and $\Lambda$CDM
haloes is in the subhaloes and associated tidal streams seen in the
$\Lambda$CDM case.  Subhaloes stand out clearly in
Fig.~\ref{fig:phasespace_caustics} since their particles have short orbital
periods and undergo many caustic passages. 

In the bottom panel of Fig.~\ref{fig:phasespace_caustics} we eliminate
self-bound subhaloes to show only particles belonging to the main halo.  Here
tidal streams corresponding to disrupted subhaloes can be traced very
well. Material from such disrupted subhaloes has a different caustic count
distribution than other main halo material, and so stands out in these
phase-space plots.  In Fig.~\ref{fig:phasespace_caustics_zoom} we show two
zooms into the phase-space structure around the most prominent subhalo of
Fig.~\ref{fig:phasespace_caustics}. These show the structure of the tidal
streams in more detail.  We note that these tidal streams are {\it not}
equivalent to the fine-grained streams we will discuss below. As we will see,
the latter are not visible in such phase-space plots because of their large
number and their low densities.

\begin{figure}
\center{
\includegraphics[width=0.45\textwidth]{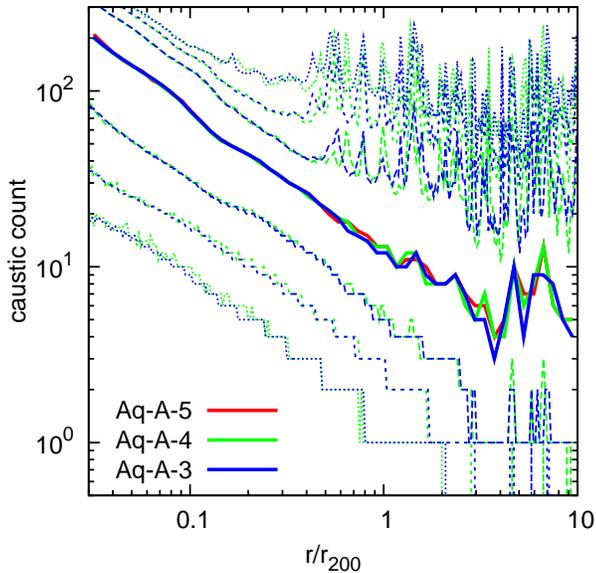}
}
\caption{Median number of caustic passages (solid thick lines) in Aq-A as a
  function of radius. In the outer regions subhaloes show up as peaks. The
  convergence between the different resolutions is very good. 
    Thin lines show the upper and lower $25\%$, $5\%$ and $1\%$ quantiles of
    the caustic count distribution in each radial bin for Aq-A-3 and
    Aq-A-4. The agreement between the distributions is excellent, at all radii
    except for some shifts at high counts in the outer regions which reflect
    small offsets in the positions of substructures. Clearly, caustic
    identification is stable against discreteness effects at these
    resolutions.}
\label{fig:caustic_passages} 
\end{figure}
\begin{figure}
\center{
\includegraphics[width=0.45\textwidth]{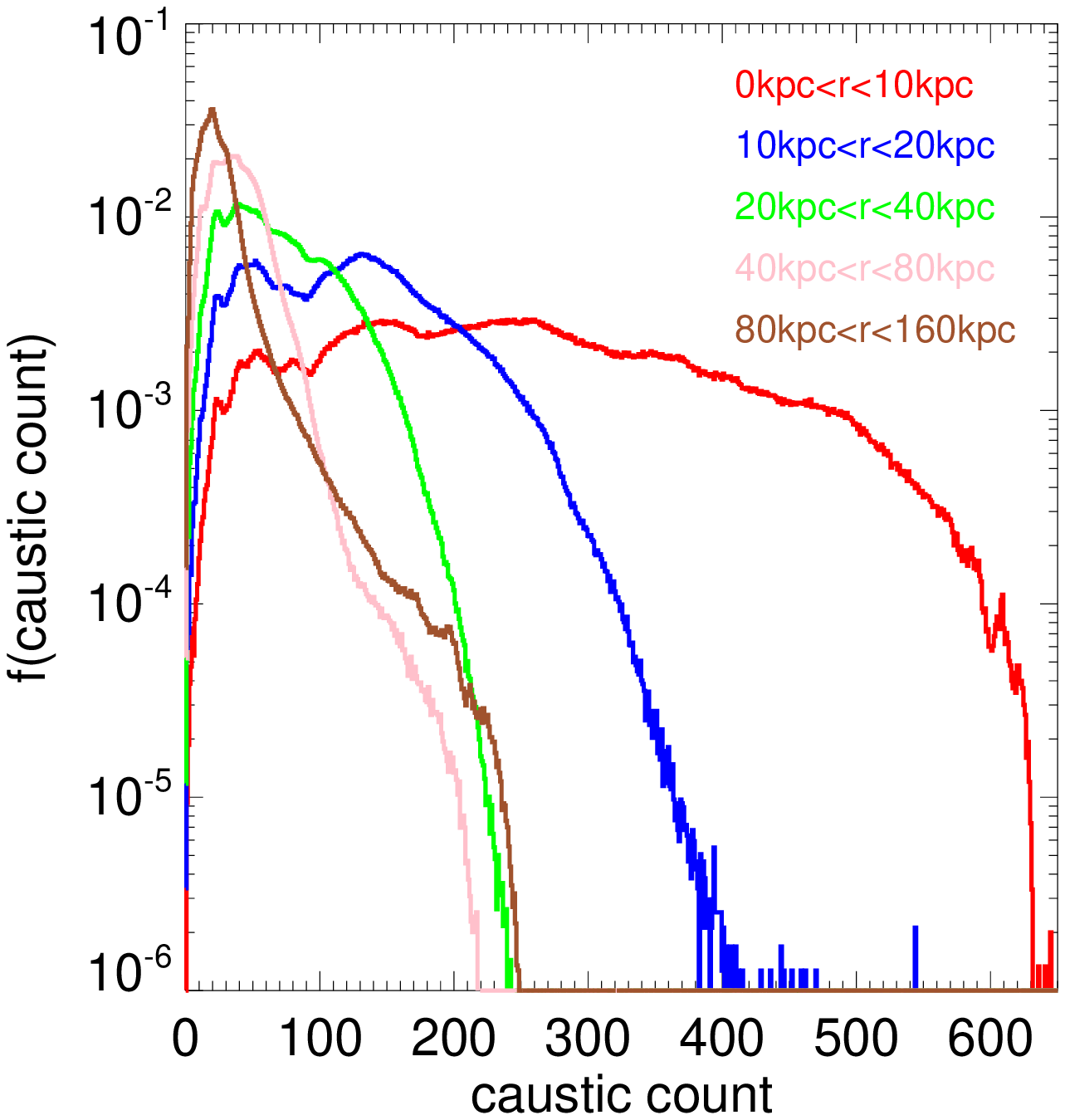}
\includegraphics[width=0.45\textwidth]{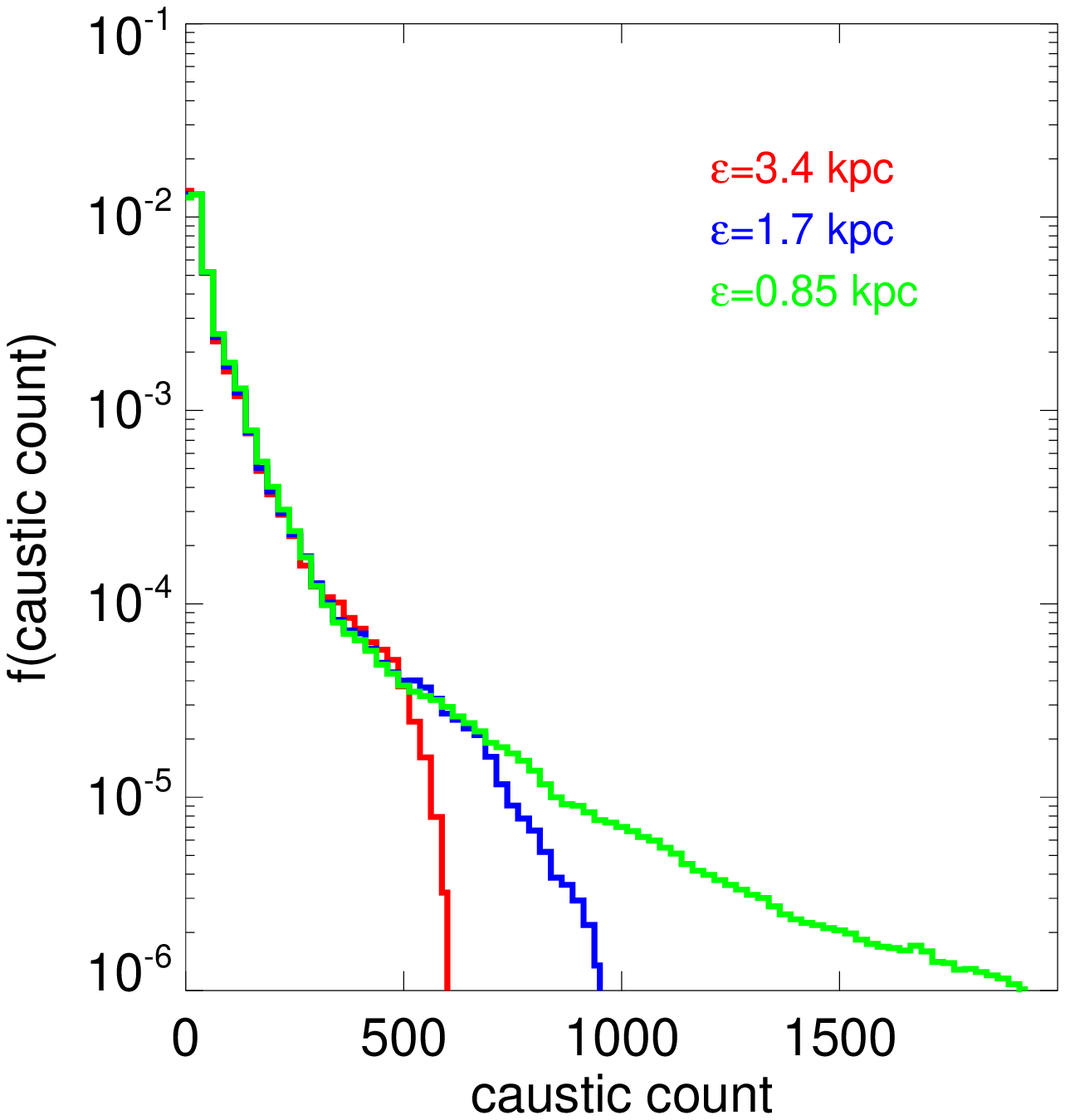}
}
\caption{Top panel: Distribution of number of caustic passages at
  $z=0$ for particles in a set of nested spherical shells in Aq-A-3
  for our standard softening of $3.4$~kpc. As expected, higher caustic
  counts are found in the inner regions. Bottom panel: Caustic count
  distributions within 160~kpc for Aq-A-3 resimulations with various
  softenings.  The smaller the softening length, the longer the tail
  of high-count particles. The particles affected are almost all in
  the innermost region of the the main halo. Note that aside from this
  tail, the histograms are almost independent of softening.}
\label{fig:caustic_hist_rad} 
\end{figure}
\begin{figure}
\center{
\includegraphics[width=0.45\textwidth]{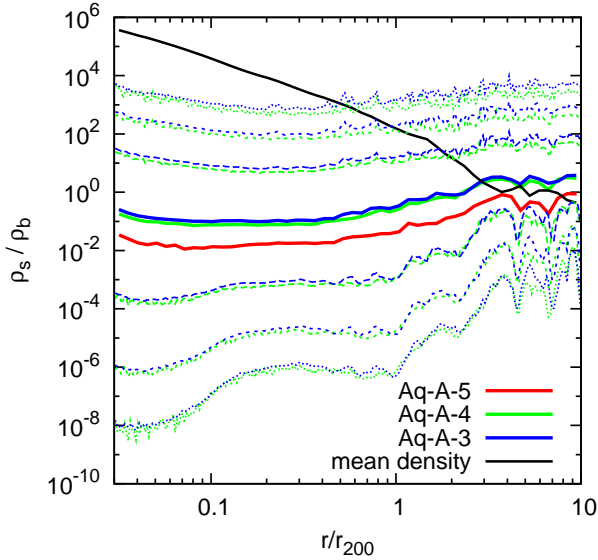}
}
\caption{The distribution of stream density in units of the cosmic mean
  density ($\rho_b$) as a function of radius within halo
  Aq-A. Continuous coloured curves give the median stream density of particles
  as a function of their distance from halo centre for Aq-A-3 (blue), Aq-A-4
  (green) and Aq-A-5 (red). Dashed and dotted curves give the
    0.5, 2.5, 10, 90, 97.5 and 99.5\% points of the distribution of stream
    density at each radius for Aq-A-3 (blue) and Aq-A-4 (green). The stream
    density distributions of the two higher resolution simulations agree well,
    but Aq-A-5 gives somewhat lower stream densities (for all percentiles, not
    just the median shown) as a result of its higher discreteness noise, We
    conclude that the resolution of Aq-A-4 is sufficient to suppress 2-body
    relaxation and other discreteness effects to an acceptably small level.
  Note that, remarkably, the median stream density is within an order of
  magnitude of the cosmic mean at all radii.  For comparison, the black solid
  curve repeats the mean mass density profile of the halo from
  Fig.~\ref{fig:mean_density}. Note that although the distribution of stream
  density spans twelve orders of magnitude in the inner halo, fewer than 1\%
  of dark matter particles are in streams with densities exceeding 1\% of the
  local mean.}
\label{fig:stream_density} 
\end{figure}
\begin{figure}
\center{
\includegraphics[width=0.45\textwidth]{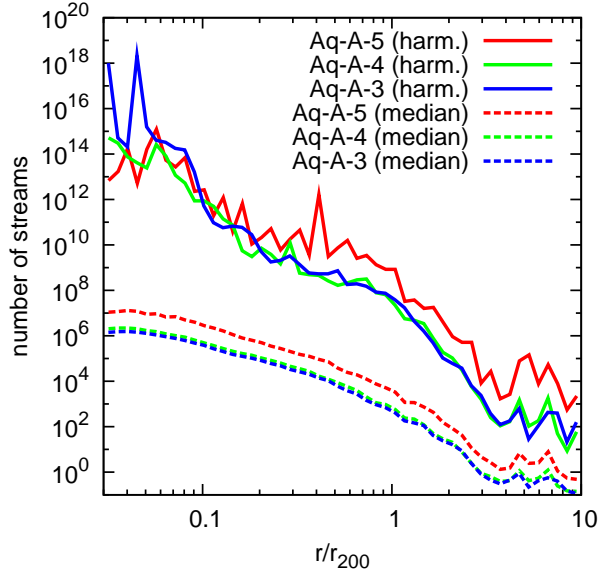}
}
\caption{The number of streams in Aq-A as a function of radius, as
  estimated from the fine-grained stream densities of individual
  simulation particles. As in Fig.~\ref{fig:stream_density}, the
  Aq-A-5 results disagree with those from the two higher resolution
  simulations, but Aq-A-3 and Aq-A-4 agree reasonably well.  Stream
  numbers are estimated by dividing the mean density at each radius by
  an estimate of the characteristic local density of individual
  streams.  For this estimate we use both the harmonic mean (solid
  lines) and the median (dashed lines) stream density for particles in
  each radial shell. The harmonic mean is very sensitive to the low
  density tail of the stream density distribution, but the agreement
  between Aq-A-3 and Aq-A-4 remains good except at the smallest
  radii. }
\label{fig:stream_number} 
\end{figure}

Caustic counts can be used to study how dynamical mixing is related to
particle location. To this end, we filter the particle distribution by
number of caustic passages and plot the result in three orthogonal
projections in Fig.~\ref{fig:caustic_nums}. The top row of this figure
shows that particles that have passed no caustic are almost
homogeneously distributed, with no clear structure.  Particles that
have passed exactly one caustic already delineate the large-scale
structure of the density field. The main filament passing through the
primary halo is visible, as is the halo itself and additional smaller
filaments pointing towards it. The halo is even more dominant
in the distribution of particles that have passed two or ten
caustics. The bottom row of Fig.~\ref{fig:caustic_nums} shows
particles that have passed at least $50$ caustics. These are found
only near the centres of the main halo and of the more massive
subhaloes. These are the densest regions with the shortest orbital
times. Typical values of the caustic count in various regions thus
indicate their level of dynamically mixing. Large values correspond to
well-mixed regions whereas small numbers correspond to dynamically
``young'' regions. Particles that have passed no caustic are still in
the quasilinear phase of structure growth.

We can compress the caustic count information in a radial profile. This
is shown in Fig.~\ref{fig:caustic_passages}.  Profiles of the median caustic
count for Aq-A-3,4 and 5 are plotted as solid curves. As already demonstrated
in \cite{2008MNRAS.385..236V}, the numerical identification of caustics is
very robust, and as a result the number of caustic passages is little affected
by numerical noise.  This is why the median profile is almost independent of
resolution in Fig.~\ref{fig:caustic_passages}, with remarkably good agreement
in the inner halo. In the outer regions subhaloes show up as peaks in the
caustic count and small shifts in their position between the different
simulations show up as apparent noise. For the highest resolution simulation,
Aq-A-3, we also plot the upper and lower 1, 5 and 25\% points of the count
distribution at each radius. These parallel the distribution of the median
count and span an order of magnitude at each radius.  The median profile can
be compared to that given by \cite{2009MNRAS.400.2174V} for an isolated halo
growing from self-similar initial conditions. At a given fraction of the
virial radius, the typical number of caustic passages is only a few times
larger in the more complex $\Lambda$CDM case.

The increase in caustic count towards halo centre can also be seen in
Fig.~\ref{fig:caustic_hist_rad} (top panel), where we present count
histograms for particles in a set of nested spherical shells. The
shift of the distributions towards lower counts with increasing radius
is very obvious, and within about 30~kpc this is accompanied by a
suppression of the high-count tail. At larger radii, however, the
distributions always extends out to about 200 counts. This tail is
contributed by particles from present or recently disrupted subhaloes
(see below). In the outer halo, the majority of particles have
nevertheless passed relatively few caustics. As noted above,
simulations using the GDE technique require significantly more
gravitational softening than standard N-body simulations in order to
limit discreteness noise in the tidal field which otherwise leads to
an unphysical, nearly exponential decay in stream density. Caustic
identification is, however, relatively stable against such effects, so
when studying caustic counts it is possible to reduce the
softening. We show the effects of this in the bottom panel of
Fig.~\ref{fig:caustic_hist_rad} which compares the caustic count
distribution within 0.16~Mpc in our standard resimulation with that in
two additional resimulations with two and four times smaller
softening. Smaller softening results in better resolution of the
innermost regions both of the main halo and of subhaloes, and thus to
shorter dynamical times and larger caustic counts for particles
orbiting in these regions. This is evident as an extension of the
high-count tail with decreasing softening. Notice, however that the
region affected is more than two orders of magnitude below the peak of
the distribution. The shift towards higher counts only affects a few
percent of particles that pass close to halo centre, and the
distribution is essentially unaffected below a count of about 400.

\subsection{The distribution of fine-grained stream density}

In the standard CDM paradigm, nonlinear evolution in the dark matter
distribution can be viewed as the continual stretching, folding and wrapping
of an almost 3-dimensional ``phase-sheet'' which initially fills configuration
space almost uniformly and is confined near the origin in velocity space
\citep[see, for example][]{2009MNRAS.392..281W}. Caustics are one generic
prediction of such evolution. Another is that the phase-space structure near a
typical point within a dark matter halo should consist of a superposition of
streams, each of which has extremely small velocity dispersion and a spatial
density and mean velocity which vary smoothly with position. If the number of
streams at some point, for example at the position of the Sun, is relatively
low, the resulting discreteness in the velocity distribution could give rise
to measurable effects in an energy-sensitive detector of the kind used in many
dark matter experiments. To assess this possibility, we need to estimate the
number and density distributions of streams at each radius.  Our GDE formalism
makes this possible by providing a value for the density of the fine-grained
stream associated with each simulation particle.  The set of stream densities
corresponding to particles within some spherical shell is thus a
(mass-weighted) Monte Carlo sampling of the stream densities at that radius.

In Fig.~\ref{fig:stream_density} we show the distribution of fine-grained
stream density as a function of radius in halo Aq-A.  
  Continuous curves give radial profiles at $z=0$ for the median stream
  density of particles at three numerical resolutions, while broken curves
  compare the tails of the distributions for the two highest resolutions.
  Agreement is excellent at all radii and throughout the distribution for
  Aq-A-3 and Aq-A-4, but the lower resolution simulation Aq-A-5 gives lower
  stream densities at all radii (and at all points in the distribution,
  although we do not show this in order to avoid confusing the plot). This is
  a consequence of discreteness noise in the tidal tensor which affects
  integration of the GDE. We note that Fig.~\ref{fig:caustic_passages} and
  Fig.~\ref{fig:stream_density} demonstrate that the caustic count and stream
  density distributions are independent of particle number at our two highest
  resolutions. Thus, discreteness effects, in particular 2-body relaxation,
  have no significant influence on the evolution.  In the Appendix we show
  that while the caustic count distribution is, in addition, independent of
  the assumed gravitational softening, the same is not true for the stream
  density distribution. This is because softening influences the tidal forces
  on particles which pass within one or two softening lengths of the centre of
  a halo or subhalo, and so influences the evolution of their stream
  density. As we discuss in the Appendix, this affects a surprisingly large
  fraction of halo particles at some point during their evolution, skewing the
  stream density distribution out to quite large radii.  For our standard
  softening length of 3.4~kpc, the effects are similar in amplitude to those
  expected from our neglect of the baryonic component, so we do not attempt to
  correct for them.

A remarkable result from Fig.~\ref{fig:stream_density} is that the
median stream density depends very little on radius, and is within an
order of magnitude of the cosmic mean density at all radii. We found a
very similar result in \cite{2009MNRAS.400.2174V} for
collapse from self-similar and spherically symmetric initial
conditions, so it is likely to apply quite generally to collisionless,
nonlinear collapse from a smooth and near-uniform early state. It
implies that the dark matter distribution from the immediate
neighborhood of a randomly chosen point in the early universe is
almost equally likely to be compressed or diluted relative to the
cosmic mean by subsequent nonlinear evolution, and furthermore that
whether it is compressed or diluted is almost independent of its
final distance from halo centre.

\begin{figure}
\center{
\includegraphics[width=0.45\textwidth]{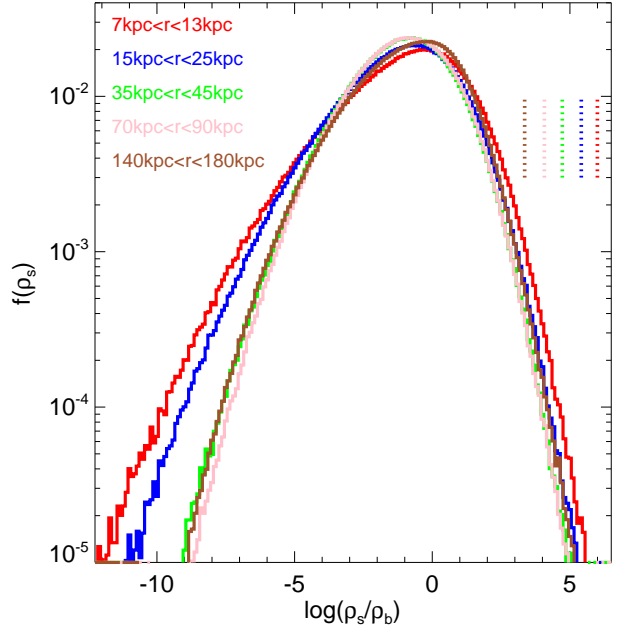}
}
\caption{Histograms of fine-grained stream density at $z=0$ for Aq-A-3
  particles in set of spherical shells. Labels with the colour of each
  histogram give the radial range of the corresponding shell, while
  vertical dotted lines correspond to the mean halo density within
  that shell.  All histograms are normalised to integrate to unity.}
\label{fig:stream_hist_rad} 
\end{figure}
\begin{figure}
\center{
\includegraphics[width=0.45\textwidth]{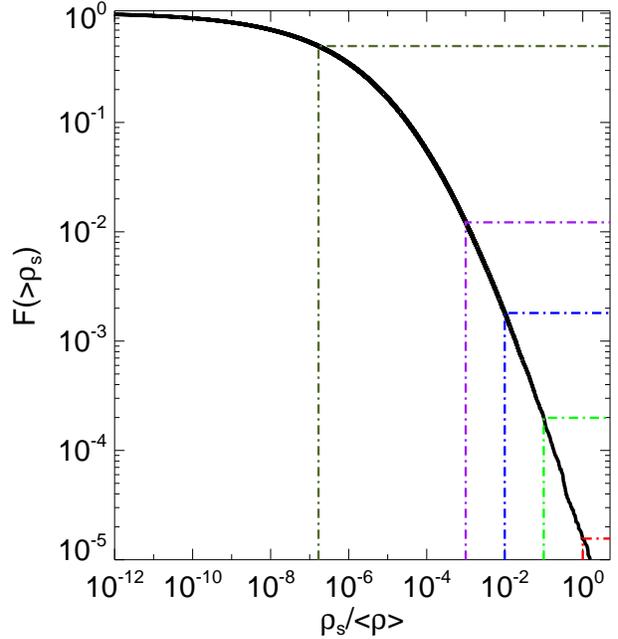}
}
\caption{The fraction of particles in Aq-A-3 at $z=0$ with halocentric radii
  in the range $7$ to $13$~kpc which have fine-grained stream density
  exceeding $\rho_s$ is plotted as a function of
  $\rho_s/\langle\rho\rangle$, where $\langle \rho \rangle$ is the mean halo
  density at $10$~kpc. Half of all particles are in streams with 
  $\rho_s > 10^{-7}\langle\rho\rangle$. See the text for a discussion of the
  other characteristic points marked.}
\label{fig:F_rho} 
\end{figure}
\begin{figure*}
\center{
\includegraphics[width=1.0\textwidth]{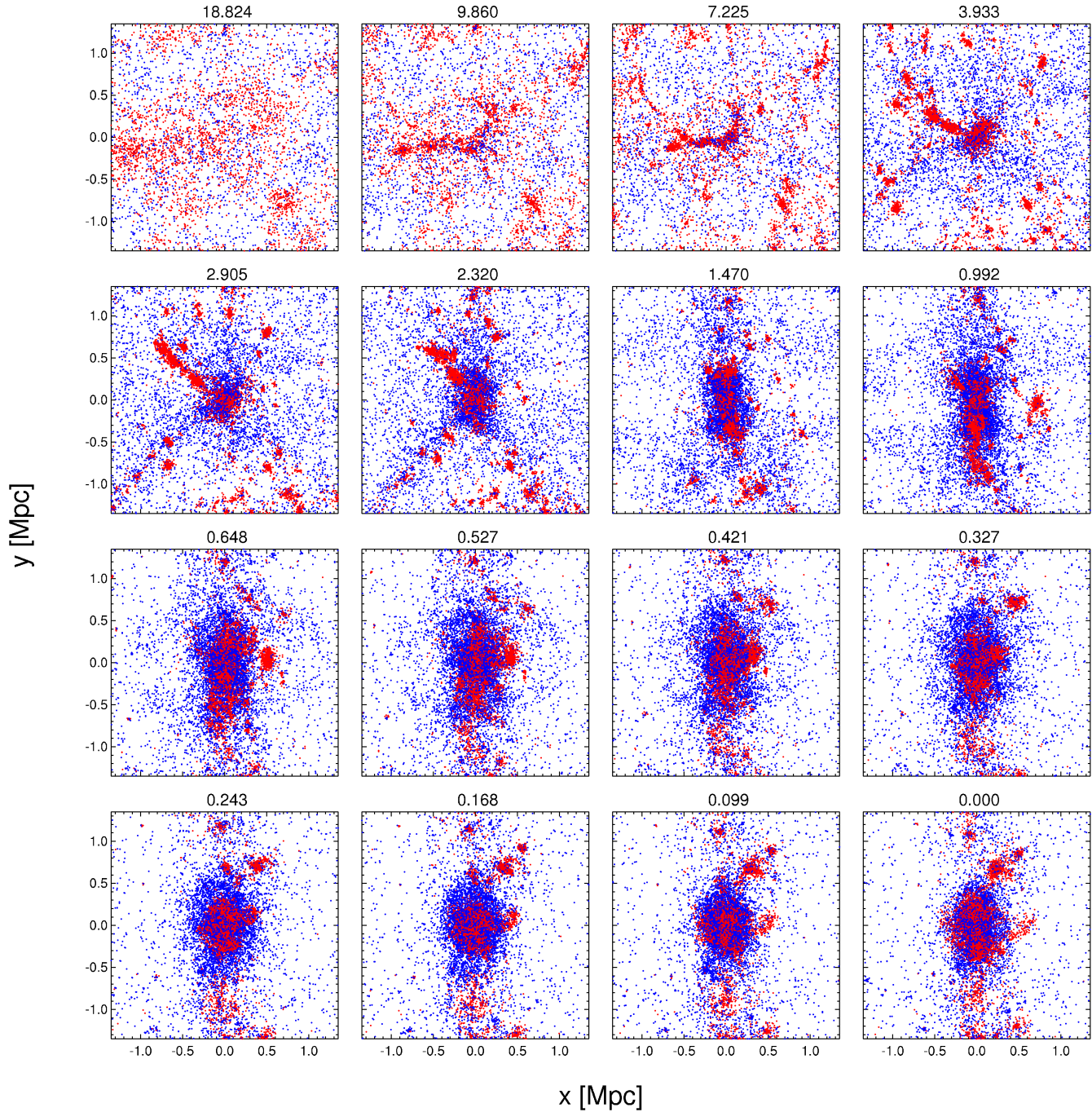}
}
\caption{The panels show a series of snapshots from the evolution of Aq-A-5.
  The redshift is indicated above each panel. Red and blue points refer to the
  same particles in each panel and are plotted in random order so that the
  probability of a random pixel being red or blue in a saturated region is
  proportional to the number of particles of each colour in that pixel.  The
  particles were selected by separating the full $z=0$ particle distribution
  into a series of thin spherical shells centred on the main halo potential
  centre, and then choosing in each shell the 1\% tails with the highest
  (blue) and lowest (red) stream density. Blue and red particles are thus
  equal in number and at $z=0$ each population is distributed in distance from
  halo centre in the same way as the particle distribution as a whole.  Only
  particles within $10~r_{\rm 200}$ at $z=0$ were considered. The comoving
  cubic region plotted lies fully within this particle set at all times and is
  centred on the centre of mass of the 200 particles which were most bound at
  $z=0$. Clearly, low $z=0$ stream density particles typically belong to
  collapsed structures at early times, whereas high stream density particles
  were generally part of no structure before they were accreted smoothly onto
  the main halo at relatively late times.}
\label{fig:evolve_decomp_s} 
\end{figure*}
\begin{figure*}
\center{
\includegraphics[width=1.0\textwidth]{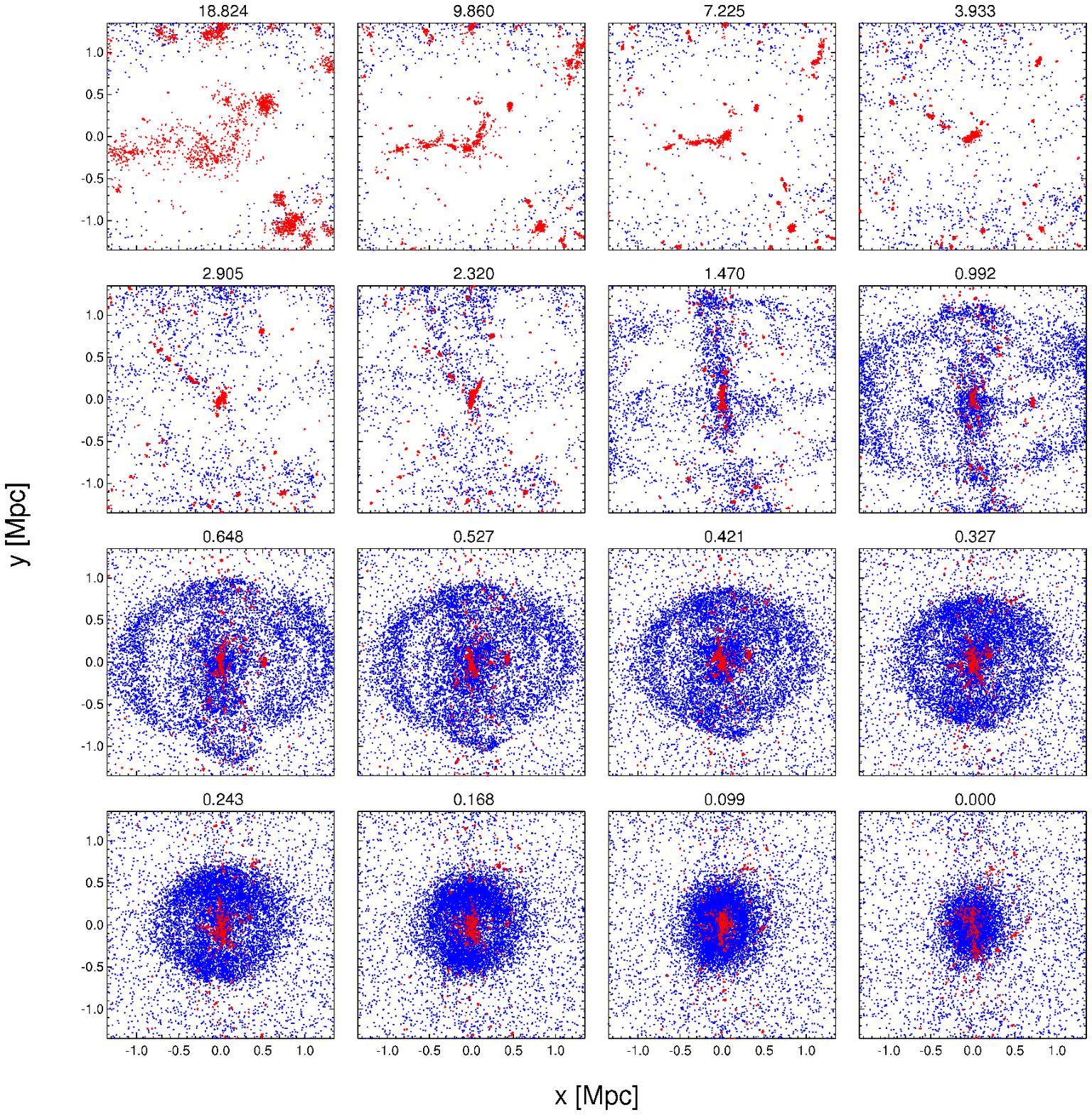}
}
\caption{This figure was constructed in exactly the same way as
  Fig.~\ref{fig:evolve_decomp_s} and shows the same number of red and blue
  particles at the same redshifts and within the same spatial regions. The
  difference is that particles were here selected on the basis of the number
  of caustics they have passed by $z=0$, rather than by their present-day
  stream density. Thus, for the same set of thin spherical shells as before,
  red particles are the 1\% within each $z=0$ shell with the highest caustic
  count, and the blue particles are the 1\% with the lowest caustic count
  (random sampled if necessary among particles with equal count to obtain
  exactly the correct number). Thus the number of red and blue points in the
  lower right panel of this figure and of Fig.~\ref{fig:evolve_decomp_s}
  is the same, and they have exactly the same distribution in halocentric
  radius. The small apparent number of red points in most panels of this
  figure is due to the fact that high $z=0$ caustic count particles are
  strongly concentrated to the centres of collapsed structures at all times.
  The relatively small number of blue points at the earliest redshifts in this
  figure reflects the fact that many blue particles are outside the region
  plotted at early times.  The qualitative behaviour of the distributions in
  this figure and in Fig.~\ref{fig:evolve_decomp_s} is similar, but the
  separation into high spatial density and near-uniform distributions is much
  more marked when particles are selected by caustic count rather than by
  stream density. In addition, low caustic count particles are accreted
  onto the final halo in a shell-like pattern which is a clear reflection of
  the structure seen in spherical infall models of halo formation.}
\label{fig:evolve_decomp_c} 
\end{figure*}

The thin dashed and dotted lines in Fig.~\ref{fig:stream_density} show the
0.5, 2.5, 10, 90, 97.5 and 99.5\% points of the stream-density distribution as
a function of distance from the centre of our Aq-A-3 and Aq-A-4
  simulations. This distribution is very broad, spanning almost 12 orders of
magnitude near halo centre. Despite this, within $0.5 r_{200}$ even the upper
0.5\% tail of stream densities lies below the mean density of the halo which
is shown as a solid black curve for comparison.  At 8~kpc, the equivalent of
the Solar radius, fewer than 1\% of all dark matter particles are part of a
fine-grained stream with density exceeding 1\% of the local mean halo density.
This is a first indication that fine-grained streams are unlikely to influence
direct detection experiments strongly.

The total number of streams at a typical point in a radial shell can
be estimated from the stream density distribution as the ratio of the
mean halo density in the shell to the harmonic mean of the stream
densities of the particles it contains. This number is dominated by
the extended low-density tail of the stream-density distribution; a
very large number of very low-density streams is predicted in the
inner regions. A more relevant measure for dark matter detection can
be obtained from the median stream density of the particles. Half of
all events in a dark matter detector will come from streams with
densities exceeding this value, and so will come from a number of
streams somewhat less than half the ratio of the mean halo density
to this median value.

In Fig.~\ref{fig:stream_number} we show the results of such
calculations for our resimulations of Aq-A at various
resolutions. Solid lines show the total number of streams at each
radius estimated from the harmonic mean stream density, whereas dashed
lines show the number of ``massive'' streams estimated from the median
individual stream density. The two estimates differ dramatically,
particularly in the inner halo, reflecting the very broad distribution
of stream densities at each radius, and in particular the presence
of simulation particles with very low stream densities.  As in
Fig.\ref{fig:stream_density}, convergence is somewhat poorer than for
the caustic count statistics we looked at above, but the results for
Aq-A-3 and Aq-A-4 nevertheless agree remarkably well. This is
particularly notable for the solid lines, given the sensitivity of the
harmonic mean to the low-density tail of the distribution. From
Fig.~\ref{fig:stream_number} we estimate the total number of streams
at halocentric radii near that of the Sun to be around $10^{14}$ and
the number of ``massive'' streams to be about $10^6$. Clearly CDM
haloes are predicted to be very well mixed in their inner regions, and
the velocity distribution near the Sun should appear very smooth.
The very large number of streams predicted near the Sun's position can be
understood from the analysis in \cite{2008MNRAS.385..236V}, which shows that
stream density should fall inversely as the number of caustic passages in a
one-dimensional system, but as the cube of this number for regular orbits in a
three-dimensional system, and even faster for chaotic orbits. Given that
typical particles near the Sun have passed a few hundred caustics, typical
stream densities are predicted to be $\sim 10^{-7}\rho_b$. The local mass
overdensity at the solar position is $\sim 10^5$, so one predicts $\sim
10^{12}$ fine-grained streams passing through our vicinity. Additional scatter
is introduced by the fact that a significant fraction of the particles are
``pre-processed'' in smaller mass systems before they fall into the main halo.

We show the distribution of stream densities in a different form in
Fig.~\ref{fig:stream_hist_rad}. For a series of spherical shells with mean
radii increasing by factors of two, we have made histograms of the
fine-grained stream densities of Aq-A-3 particles, normalising them to unity
so that their shapes can be compared.  Beyond 30~kpc these histograms are all
quite similar, and resemble slightly skewed log-normal distributions. At the
bottom of our plot, three orders of magnitude below peak, they span 14 orders
of magnitude in stream density. At smaller radii, the low-stream-density tail
becomes more extended, the peak of the distribution shifts very little, and
the shape of the high-stream-density tail is unchanged.  The latter is
determined by the behaviour near caustics. As we will see below the {\it
  maximum} densities at caustics are predicted to be in the range explored by
this high-mass tail. It is notable that within 20~kpc these tails do not
extend up to the local mean density of the halo.

The implications of these distributions for direct detection experiments on
Earth are most easily drawn from Fig.~\ref{fig:F_rho}, which shows the
cumulative distribution of fine-grained stream density for Aq-A-3 particles
with $7~{\rm kpc} < r < 13~{\rm kpc}$ at $z=0$.  Specifically, we plot the
fraction of particles with stream density exceeding $\rho_s$ against
$\rho_s/\langle\rho\rangle$, where $\langle\rho\rangle$ is the mean halo
density at 10~kpc. The fraction of particles with $\rho_s >
\langle\rho\rangle$ is about $2\times 10^{-5}$, so the probability that a
single stream dominates the signal in a direct detection experiment (i.e. that
the Earth lies sufficiently close to a sufficiently strong caustic) is about
$2\times 10^{-5}$.  The fraction of particles with $\rho_s > 0.1
\langle\rho\rangle$ is about $2\times 10^{-4}$, hence the probability to see a
single stream containing 10\% of the signal is about 0.002. Similarly, the
fraction of particles with $\rho_s > 0.01 \langle\rho\rangle$ is about
$2\times 10^{-3}$, so a single stream containing 1\% of the signal will be
seen with probability 20\%. Finally, the fraction of particles with $\rho_s >
0.001 \langle\rho\rangle$ is about $10^{-2}$; this means that at a typical
``Earth'' location there will be a few streams which individually contribute
more than 0.1\% of the local dark matter density.  Thus an experiment which
registers 1000 true dark matter ``events'' should get a few duplicates,
i.e. events coming from the same fine-grained stream and so having (very
nearly) the same velocity. If the dark matter is made of axions then a
resonant detector should find an energy spectrum where a few tenths of a
percent of the total energy density is concentrated in a few very narrow
``spectral lines''.

\begin{figure}
\center{
\includegraphics[width=0.45\textwidth]{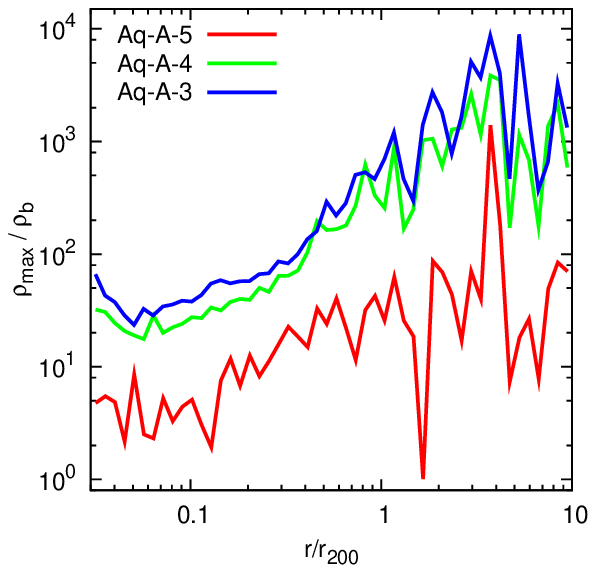}
\includegraphics[width=0.45\textwidth]{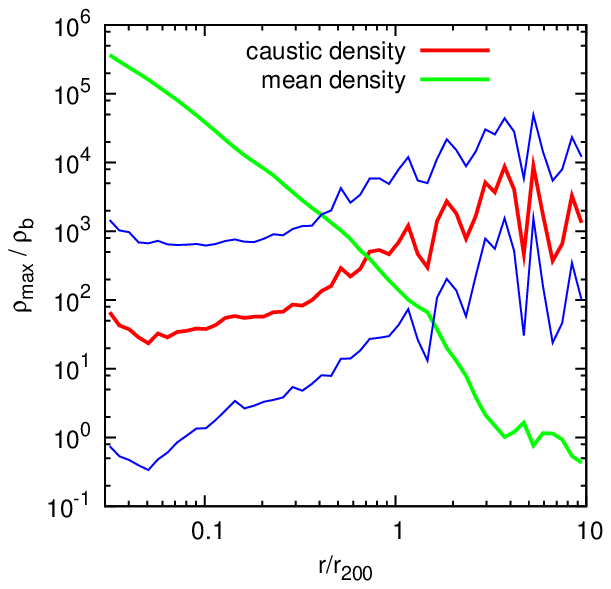}
}
\caption{Top panel: Radial profile of the median value of peak caustic density
  for simulations at three different resolutions. The resolution of Aq-A-5 is
  too low to get converged results, but the results from Aq-A-4 and Aq-A-3
  agree quite well. Bottom panel: The mean halo density profile of Aq-A-3 is
  compared to the median and the quartiles of the peak density of caustics as
  a function of radius. Beyond the virial radius, typical caustics have peak
  densities which exceed the local mean density. Caustics in the inner halo
  have very low contrast, however.}
\label{fig:max_caustic_density} 
\end{figure}

\subsection{Origin of the tails of the stream density and caustic count distributions}

Given the very broad range of stream densities and caustic counts present at
each point within our haloes, it is interesting to investigate how these
properties are related to the dynamical history of individual particles.
We begin by studying stream density.

As we saw above, the ``typical'' stream density of particles (specifically,
the local median value) is almost independent of local mean halo density, thus
of local dynamical time.  Lower stream density corresponds to greater
stretching of the initial phase-sheet and so, one might think, to more
effective dynamical mixing. An anticorrelation of fine-grained phase density
with local orbital time (i.e. with radius) is, however, at most marginally
evident in the stream density distributions of Fig.~\ref{fig:stream_density}
and then only in their low-stream-density tails. To a first approximation
there is no correlation between stream density distribution and radius.  The
overall mean density profile of the halo is built up by having more streams of
every density at smaller radii. Thus the relation of stream density to
dynamical mixing is far from evident.

\begin{figure}
\center{
\includegraphics[width=0.45\textwidth]{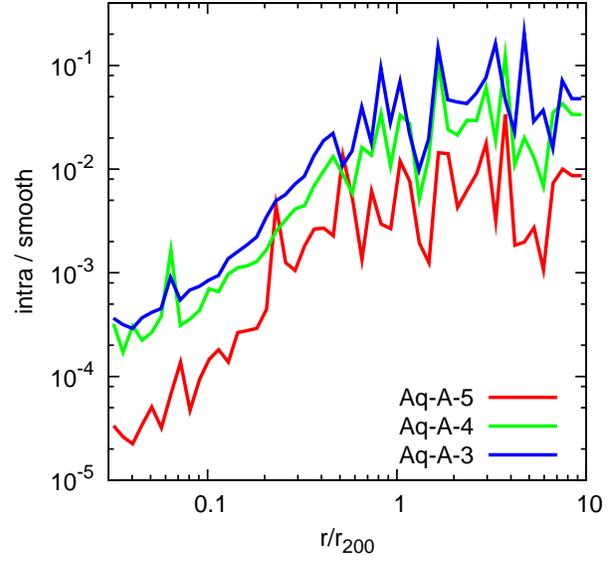}
}
\caption{Local ratio of the spherically averaged intra-stream annihilation
  rate to the spherically averaged smoothed annihilation rate as a function of
  radius. The smoothed annihilation rate is calculated from a SPH-based
  density estimate using 64 neighbours. The intra-stream annihilation rate is
  dominated by contributions around caustics. The standard ``boost factor''
  due to unbound small-scale structure is one plus the quantity plotted on the
  vertical axis. Clearly, caustics play essentially no role in enhancing the
  dark matter annihilation luminosity of the inner halo.  Around the virial
  radius they contribute about $10\%$ to the annihilation signal.}
\label{fig:boostprofile} 
\end{figure}

We explore this issue in Fig.~\ref{fig:evolve_decomp_s}.  Red and blue points
here highlight the positions at a series of redshifts of particles which at
$z=0$ lie within $10~r_{\rm 200}$ of halo centre. We divided this region into
a set of thin nested spherical shells and then for each shell we identified
particles in the upper and lower 1\% tails of the stream density
distribution. The particles with the highest $z=0$ stream densities (which are
typically close to caustics) are plotted blue in each panel while those with
the lowest stream densities (which are typically in collapsed regions and far
from caustics) are plotted red. Thus at $z=0$ there are the same number of red
and blue points in the plot, and both populations are distributed in
halocentric radius exactly as the mass distribution as a whole. The spatial
distributions of the two populations differ markedly, however, both at $z=0$
and at all earlier times. Most of the red particles are part of clumps already
at early times, while most of the blue particles either appear diffuse or are
distributed relatively smoothly through the main halo at all times.  There is
thus indeed a correlation between stream density and dynamical history,
despite the lack of correlation between stream density and halocentric
distance in Fig.~\ref{fig:stream_density}.

The number of caustic passages is a clearer measure of dynamical mixing,
because it is closely related to the number of orbits a particle has executed
over its entire dynamical history. We examine this using
Fig.~\ref{fig:evolve_decomp_c} which is constructed in exactly the same way as
Fig.~\ref{fig:evolve_decomp_s} except that the particles in each $z=0$
spherical shell were ranked by caustic count rather than by stream density.
Qualitatively the two series of plots show some similarities, but the
differences between the red and blue populations are substantially more marked
in the caustic count case. The red points are almost all very close to the
centres of collapsed clumps at all times, whereas the blue population is
always diffuse except at low redshift when some of it is accreted onto the
main halo. The evolution of the blue particles in
Fig.~\ref{fig:evolve_decomp_c} shows an interesting shell-like structure which
can be understood with reference to a spherical infall model. Many of the blue
particles have yet to pass a caustic at $z=0$, and so are on their first
passage through the halo.  Such particles come from a narrow range of
Lagrangian radii in the initial conditions, hence the apparent shell. The fact
that the caustic count is a much better indicator of the overall level of
dynamical mixing than the stream density is easily understood; the stream
density varies strongly along the trajectory of each individual particle
reaching high values each time it passes a caustic, whereas the caustic count
increases monotonically with the number of orbits completed.

\subsection{Caustics and dark matter annihilation}
 
The number of fine-grained streams passing through the Solar System is of
interest for the search for dark matter using laboratory devices. Caustics, on
the other hand, could be of interest for indirect dark matter searches that
focus on the annihilation products of dark matter particles.  Annihilation
explanations for many of the apparent ``anomalies'' in indirect detection
signals, for example the rise in the positron fraction detected by PAMELA
above the expectation from secondary production, require annihilation rates to
be substantially ``boosted'' above predictions based on ``standard'' particle
physics assumptions and haloes with a locally smooth dark matter density
field. The annihilation rate per unit volume is proportional to the square of
the local dark matter density and so can reach very high values in caustics.
Caustics could thus, at least in principle, provide the necessary boost.  To
quantify this with our simulations, it is important not only to identify
caustics, but also to estimate their peak densities, since emission around the
peak dominates caustic luminosity \citep[see][]{2009MNRAS.392..281W}.

In Fig.~\ref{fig:max_caustic_density} we plot median peak caustic density as a
function of radius for Aq-A. To make this plot we recorded the peak caustic
density and the radial position at which it was achieved for all particles
that pass through a caustic in a short time interval around $z=0$. We bin
these caustic passages into a set of radial shells, and then find the median
value of peak caustic density for each shell. The top panel shows results at
three different resolutions, Aq-3, Aq-A-4 and Aq-A-5. Again, it is clear that
the lower resolution of Aq-A-5 has significantly affected the results, while
Aq-A-3 and Aq-A-4 agree reasonably well. The peak caustic density calculation,
like that of the stream density, depends on the full phase-space distortion
tensor, so it is not surprising that the convergence properties of the two
measures are similar.  In the bottom panel we compare the spherically averaged
mass density profile of Aq-A-3 with the radial distribution of peak caustic
density. The thick red line repeats the median profile from the top panel,
while thin blue lines show the upper and lower quartiles of peak caustic
density as a function of radius. Notice that at each radius the distribution
of peak caustic densities is very broad, mirroring the range of fine-grained
stream densities seen in Fig.~\ref{fig:stream_density}. The peak densities of
almost all caustics in the inner part of the halo are substantially below the
local mean halo density. A quarter of all caustics have peak densities
exceeding the local mean at about $0.4 r_{200}$, but only outside $0.8r_{200}$
do more than half of the caustics have peak densities higher than the local
mean density.

The low peak densities predicted for caustics in the inner halo, suggest that
any annihilation boost will be small. This was already the case for the
isolated, smoothly growing halo studied in \cite{2009MNRAS.400.2174V}. In a
$\Lambda$CDM halo, additional small-scale structure should decrease stream
densities, and thus peak caustic densities, even further. To investigate this,
we calculate the ratio of intra-stream annihilation (both particles belonging
to the same fine-grained stream) to inter-stream annihilation (the two
particles belonging to different fine-grained streams). The former can be
calculated for each simulation particle from its GDE-estimated stream density
and can be integrated correctly through caustics using the formalism of
\cite{2009MNRAS.392..281W}. The latter can be calculated from a SPH kernel
estimate of the local smooth dark matter density.  For the boost to be
substantial the ratio of these two annihilation rates should be large.  In
Fig.~\ref{fig:boostprofile} we plot this ratio for Aq-A as a function of
radius, after averaging the rates over thin spherical shells.  Again, results
for Aq-A-3 and Aq-A-4 agree well, while those for Aq-A-5 are clearly affected
by its low resolution. This reflects the similar effects seen in earlier plots
of stream and caustic densities.  Fig.~\ref{fig:boostprofile} shows that the
contribution of caustics to the overall annihilation luminosity is very small
-- boosting is completely negligible in the inner halo, and is still small
(about a factor of 1.1) near the virial radius.  This is even lower than the
already small effects seen in the smooth halo collapse simulation presented by
\cite{2009MNRAS.400.2174V}.  Since an annihilation interpretation of the
PAMELA results requires a boost factor of 100 to 1000, caustics are clearly
far too weak to provide an explanation. Since bound subhaloes are also unable
to produce local boosts much above unity near the Sun
\citep{2008Natur.456...73S}, non-standard particle physics such as Sommerfeld
enhancement \citep[e.g.][]{1931AnP...403..257S,2004PhRvL..92c1303H,2005PhRvD..71f3528H,
2007NuPhB.787..152C,2009PhRvD..79a5014A,2009PhRvD..79h3523L} must be invoked to explain the
PAMELA data through annihilation.

\subsection{Object-to-object scatter}

In the last few subsections we discussed the fine-grained structure of Aq-A in
considerable detail and studied how it is affected by numerical resolution. It
is not, of course, clear how representative these results are, since they are
based on a single Milky Way-like halo. To quantify the scatter expected as a
result of variations in formation history, environment, etc., it is important
to check some of these properties for other haloes.  The Aquarius Project
simulated six Milky Way-mass haloes at very high resolution, and so provides
an opportunity to address this issue. The question of how representative these
particular haloes are of the full population of similar mass objects is
discussed by \cite{2010MNRAS.406..896B}. Here, we concentrate on the caustic
count profile which we showed in Fig.\ref{fig:caustic_passages} to be very
well converged in Aq-A-3, Aq-A-4 and Aq-A-5, at least within $r_{200}$ where
the ``noise'' due to substructure is small. Given this robust result, we
decided for the following comparison to rerun the other five Aquarius haloes
at resolution level 4 in order to save computation time. All resimulations
were carried out with our standard softening length of $3.4$~kpc.

\begin{figure}
\center{
\includegraphics[width=0.45\textwidth]{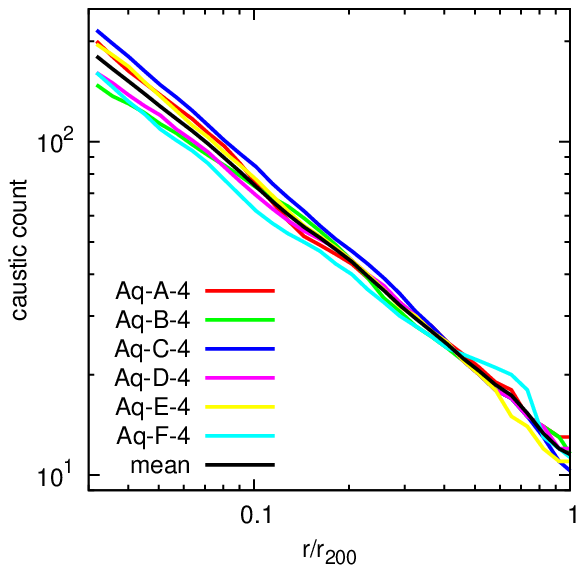}
\includegraphics[width=0.45\textwidth]{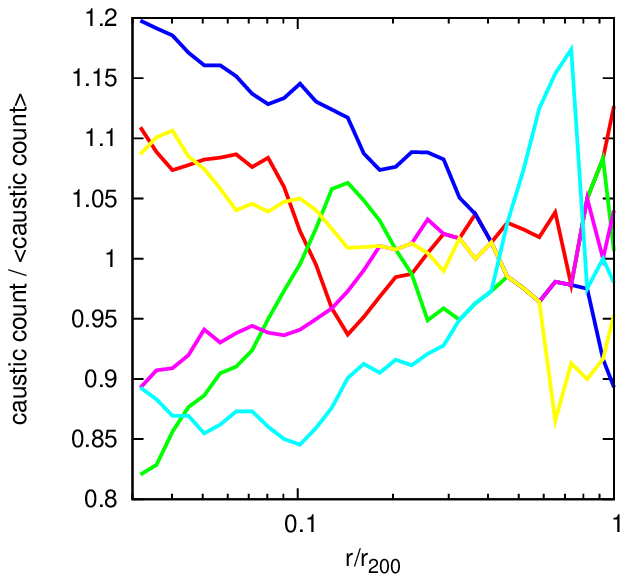}
}

\caption{Top panel: Radial profiles of median caustic count similar to that in
  Fig.\ref{fig:caustic_passages} but for all six Aquarius haloes resimulated
  at resolution level 4 with a softening of 3.4~kpc. We plot only the
  region interior to $r_{200}$ in order to exclude the substantial ``noise''
  at larger radii due to individual massive subhaloes.  Bottom panel: Ratio of
  the individual halo caustic count profiles to their mean. }
\label{fig:caustic_COM} 
\end{figure}

In the top panel of Fig.~\ref{fig:caustic_COM} we show profiles of median
caustic count as a function of radius for all six Aquarius haloes. We plot
results within $r_{200}$ only, since at larger radii the profiles are subject
to large stochastic fluctuations at the positions of massive subhaloes (see
Fig.\ref{fig:caustic_passages}). The variations between the six haloes are
relatively small and are systematic with radius as can be seen in the lower
panel of Fig.~\ref{fig:caustic_COM} where we plot the ratio of each of the
individual profiles to their mean. Differences are largest in the central
regions but still scatter by less than 20\% around the mean.  There is some
correlation of the caustic count profile with the radial mass density
profile. Haloes A and C have the most concentrated mass density profiles (see
\cite{2010MNRAS.406..896B}) and also have the highest caustic counts in the
inner regions, as might naively be expected. Halo E, on the other hand, is the
least concentrated of all the haloes and yet also has a relatively high caustic
count near the centre. Clearly, the details of halo assembly history do affect
the median number of caustic passages significantly. This is not determined
purely by the local dynamical time of the final halo.

\begin{figure}
\center{
\includegraphics[width=0.45\textwidth]{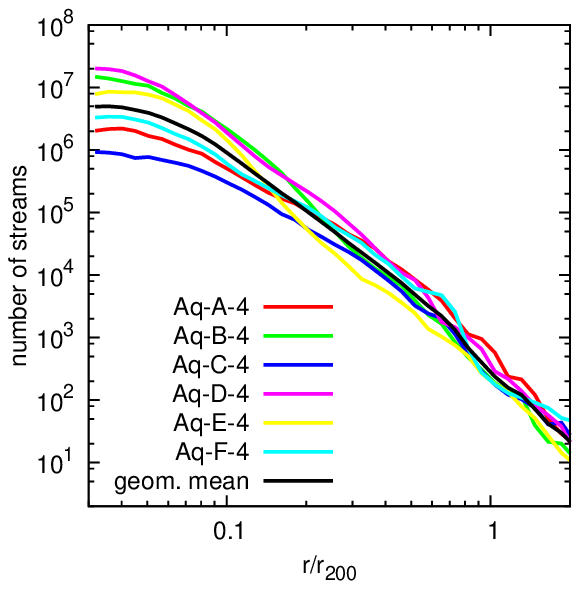}
\includegraphics[width=0.45\textwidth]{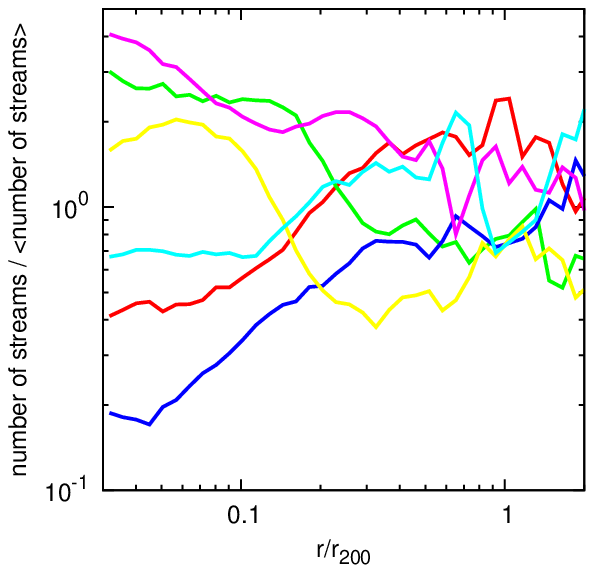}
}
\caption{Top panel: Stream number, defined as for the dashed curves of
  Fig.~\ref{fig:stream_number}, for the six Aquarius haloes at level 4
  resolution and with our fiducial softening length of $3.4$~kpc. We
  plot out to $2r_{200}$ since at larger radii these profiles are 
  affected by individual massive subhaloes. Bottom panel: Ratio of the
  individual halo stream number profiles to their (geometric) mean.}
\label{fig:stream_number_COM} 
\end{figure}

In Fig.~\ref{fig:stream_number_COM} we do the same exercise for the number of
streams. For each of our resimulated Aquarius haloes we use the median stream
density of the particles at each radius to estimate a characteristic stream
number, as in the dashed curves of Fig.~\ref{fig:stream_number} which
demonstrate that good convergence is achieved at resolution level 4. We plot
the profiles out to $2r_{200}$ in this case; at larger radii stochastic
effects due to massive subhaloes again dominate the variations. The top panel
shows the profiles themselves, while the lower panel plots the ratio of each
individual profile to their (geometric) mean. Halo-to-halo differences here
are larger than for the caustic counts, but still relatively modest given the
large dynamic range of the radial variation. It is interesting that the
ranking of haloes here is similar but not identical to that in the caustic
count profiles of Fig.~\ref{fig:caustic_COM}; the most concentrated haloes A
and C not only have the largest caustic counts in their inner regions, but also
the smallest numbers of streams. This is unexpected since a naive argument
might have led one to expect that a larger number of caustics would
correspond to a {\it larger} number of streams.

\section{Concluding remarks}

We have demonstrated how N-body simulation techniques can be extended to
follow fine-grained structure and the associated caustics during the formation
of dark matter haloes from fully general $\Lambda$CDM initial conditions. We
have shown that, for the large particle numbers of our
  standard experiments, our integrations of the geodesic deviation equation
  (GDE) produce distributions of fine-grained stream density and caustic
  structure which are independent of particle number, hence insensitive to
  discreteness effects such as two-body relaxation. This requires somewhat
  larger gravitational softenings than are used in traditional N-body
  simulations, and we show in the Appendix that while caustic count
  distributions are insensitive to the assumed softening, the same is not true
  for the stream density distributions.  We have resimulated the six Milky
Way-mass haloes of the Aquarius Project, allowing us to analyse the scatter in
fine-grained properties among haloes of similar mass.

By identifying caustic passages along its trajectory, we are able to assign a
cumulative caustic count to each simulation particle which is robust against
numerical noise and serves to indicate the extent of dynamical mixing in its
phase-space neighborhood. This caustic count provides an excellent means to
highlight subhaloes and tidal streams in $r$--$v_{r}$ phase-space plots,
because particles which became part of condensed structures at early times
complete more orbits, and so pass more caustics than particles which remain
diffuse until accreted onto the main halo.  Filtering the particle
distribution by caustic count decomposes it into interpenetrating components
which have experienced different levels of dynamical mixing. Particles that
have passed no caustic form an almost uniform subcomponent, while particles
with large caustic count (e.g. $>50$) are found only in the dense inner
regions of haloes. Particles with intermediate count outline the skeleton of
the cosmic web.

Direct dark matter detection experiments are, in principle, sensitive to the
fine-grained structure of the Milky Way's halo at the position of the Sun.  It
is especially important to know if a significant fraction of the local dark
matter density could be contributed a one or a few fine-grained streams, since
each of these would be made of particles of a single velocity with negligible
dispersion. If many streams contribute to the local density and none is
dominant, then a smooth velocity distribution can be assumed. Our simulations
show the unexpected result that the distribution of fine-grained stream
density is almost independent of radius within the virialised region of dark
matter haloes. Only the low-density tail of the distribution appears to extend
downwards at small radii. Because of the extreme breadth of the distribution,
a very large number of streams ($\sim 10^{14}$) is predicted at the Solar
position, but about half of the total local dark matter density is contributed
by the $10^6$ most massive streams. The most massive individual stream is
expected to contribute about 0.1\% of the local dark matter
density. This is potentially important for axion detection
  experiments which have extremely high energy resolution and may be able to
  detect 0.1\% of the total axion energy in a single ``spectral line''. Apart,
  from this possibility, dark matter detection experiments may safely assume
the velocity distribution of the dark matter particles to be smooth.

It has been suggested that the very high local densities associated with
caustics might significantly enhance dark matter annihilation rates in dark
matter haloes, and thus be of considerable significance for indirect detection
experiments. Indeed, for completely cold dark matter it can be shown that the
contribution from caustics is logarithmically divergent and hence dominant
\citep{2001PhRvD..64f3515H}.  This divergence is tamed by the small but finite
initial velocity dispersions expected for realistic CDM candidates, and
calculations for idealised, spherical self-similar models indicate quite
modest enhancements for standard WIMPs \citep{2006MNRAS.366.1217M}.  Our
methods allow us to identify all caustics during halo formation from fully
general $\Lambda$CDM initial conditions. We find that they are predicted to
make a substantially smaller contribution to the local annihilation rate than
in the spherical model. This is because the more complex orbital structure of
realistic haloes results in lower densities for typical fine-grained streams
and thus to lower caustic densities when these streams are folded. The
enhancement due to caustics near the Sun is predicted to be well below $0.1\%$
and so to be completely negligible.  Only in the outermost halo does the
enhancement reach 10\%. The standard N-body technique of estimating
annihilation rates from SPH estimates of local DM densities \citep[see, for
  example,][]{2008Natur.456...73S} should therefore be realistic, and such
estimates will not be significantly enhanced by caustics.  Similarly, caustics
cannot be invoked to provide the large boost factors required by annihilation
interpretations of `anomalies' in recently measured cosmic ray spectra from
the PAMELA or ATIC experiments. Given that unresolved small subhaloes also
appear insufficient to provide a substantial boost
\citep{2008Natur.456...73S}, an annihilation interpretation of these signals
is only tenable with nonstandard annihilation cross-sections.

Our comparison of results for the six different haloes of the Aquarius Project
showed that the halo-to-halo variation in the fine-grained properties we have
studied is relatively small, and thus does not have significant impact on the
applicability of our principal conclusions to the particular case of our own
Milky Way.  Thus our final conclusion must be that the fine-grained structure
of $\Lambda$CDM haloes has almost no influence on the likelihood of success
of direct or indirect dark matter detection experiments.

Although we demonstrated that two-body relaxation and other discreteness
effects have no significant influence on our conclusions, we found that our
predicted stream density distributions are strongly affected by gravitational
softening, which modifies the tidal forces on particles passing within a
softening length or two of halo centre.  This numerical effect remains a
significant uncertainty in our results. The analysis in the Appendix shows
that reducing the softening below our standard value tends to reduce the
density of streams, thus to suppress further the importance of fine-grained
structure for detection experiments.  However, the change in orbit structure
induced by our standard softening is similar in scale and amplitude to that
caused by the neglected baryonic components of the galaxies, and even the sign
of the effect of the latter is uncertain. This is clearly an area where
further work is required to reduce the remaining uncertainty in our
quantitative conclusions.

\section*{Acknowledgements}
The simulations for this paper were carried out at the Computing Centre of the
Max-Planck-Society in Garching.  MV thanks Stephane Colombi, Roya Mohayaee and
Volker Springel for helpful discussions.

\begin{appendix}
\newpage
\section{Effects of gravitational softening}

\begin{figure}
\center{
\includegraphics[width=0.435\textwidth]{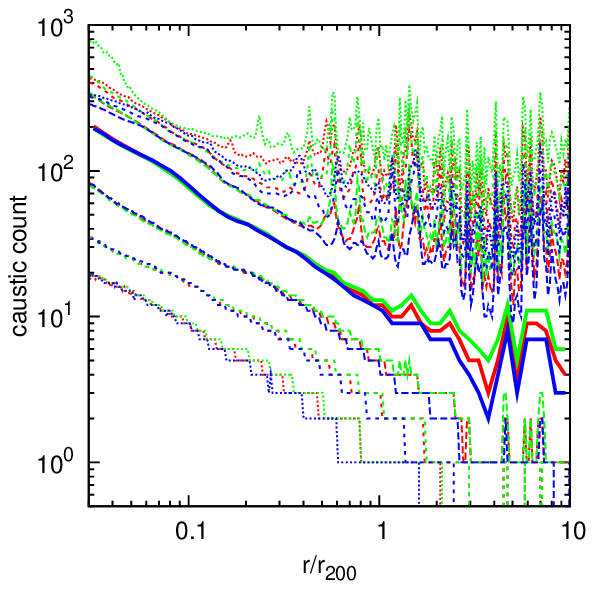}
\includegraphics[width=0.445\textwidth]{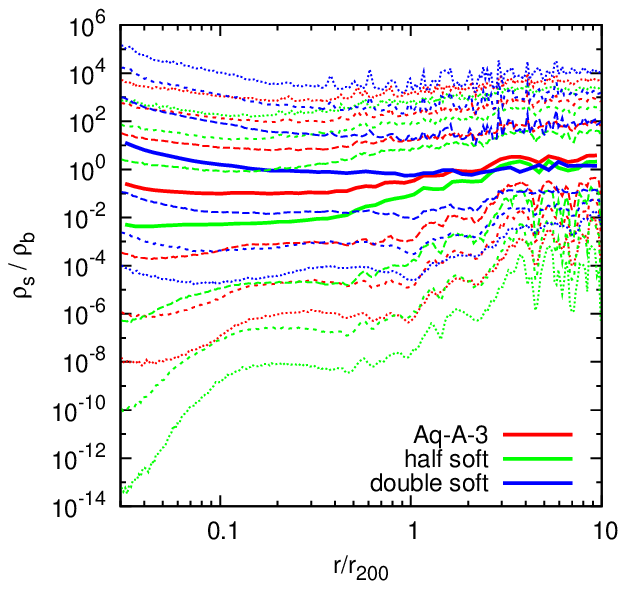}
}
\caption{Top panel: Distribution of caustic passages in Aq-A-3 as a
  function of radius and of gravitational softening. Thick lines show the
  median and thin lines the upper and lower $25\%$, $5\%$ and $1\%$ quantiles
  of the caustic count distribution at each radius.  Red lines refer to our
  standard simulation with our fiducial softening length of $\epsilon=3.4~{\rm
    kpc}$.  Green and blue lines show results for softening lengths half and
  twice this value, respectively. Clearly softening does not significantly
  affect these distributions.  Bottom panel: The distribution of stream
  density in Aq-A-3 (in units of the cosmic mean) as a function of radius and
  of gravitational softening. Continuous curves give the median, and dashed
  and dotted lines the 0.5, 2.5, 10, 90, 97.5 and 99.5\% points of the
  distribution at each radius. Red curves correspond to our fiducial softening
  length of $\epsilon=3.4~{\rm kpc}$, green and blue lines to softenings half
  and twice as big, respectively.  The stream density distribution is
  significantly affected by gravitational softening at all radii, although the
  effect gets weaker in the outer halo.}
\label{fig:profile_check} 
\end{figure}

According to Figs.~\ref{fig:caustic_passages} and \ref{fig:stream_density} the
caustic count and stream density distributions are independent of particle
number at our two highest resolutions. This demonstrates that discreteness
effects, in particular two-body relaxation, are not significantly affecting
the phase-space quantities we derive from our GDE integrations.  In this
Appendix we investigate whether these distributions are sensitive to our
assumed gravitational force softening. To facilitate this, we keep the {\it
  particle number} fixed and change the {\it softening length}. Specifically,
we resimulated Aq-A-3 using softening lengths half and
and twice our fiducial value, $\epsilon=3.4~{\rm kpc}$, but with all other
parameters held to their original values.

\begin{figure}
\center{
\includegraphics[width=0.43\textwidth]{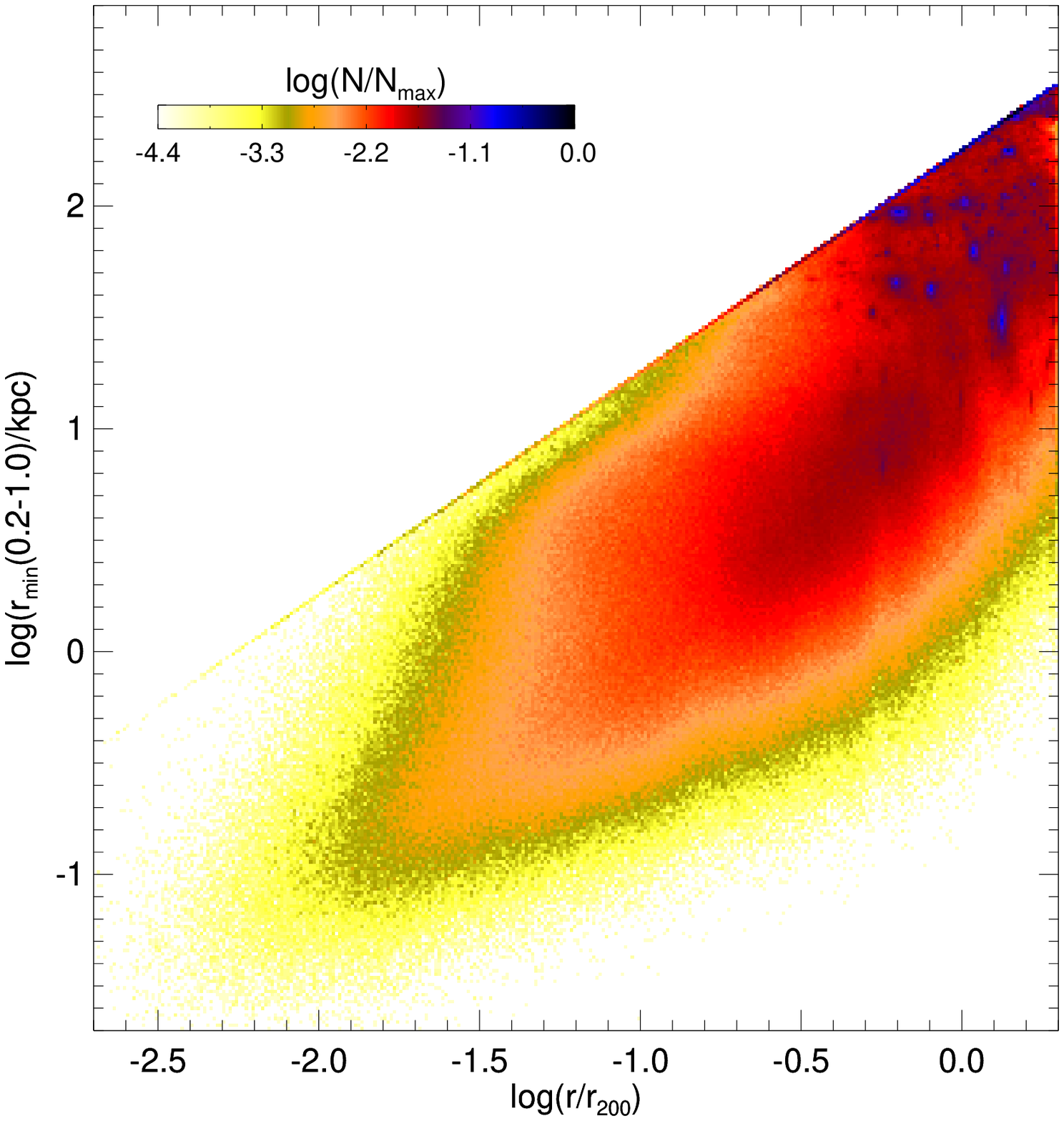}
}
\caption{ Distribution of halo particles at $z=0$ in the $r/r_{200}$ 
  versus $r_{\rm min}(0.2-1)$ plane for the Aq-A-4 simulation with half
  our fiducial softening, i.e.  $\epsilon=1.7~{\rm kpc}$. Even
  at large radius, a surprisingly large fraction of particles have
  minimal pericentric distance within one or two softening lengths.  }
\label{fig:hist_peri} 
\end{figure}

The top panel of Fig.~\ref{fig:profile_check} shows that the caustic count
distribution is independent of softening in the main halo, but that its high
count tail is affected by softening in subhaloes.  Subhaloes are effectively
small $N$ systems, and their centres are significantly better resolved with a
smaller softening length. This results in larger caustic counts at the
centre of subhaloes, which are evident as enhanced ``noise'' in the high-count
tail at larger radii for the simulation with the smallest softening.  This
effect is almost absent in the main halo. We conclude that our main results
for caustic counts are robust to changes in softening length, a conclusion
supported by the results of \cite{2008MNRAS.385..236V}.

The bottom panel of Fig.~\ref{fig:profile_check} shows a similar plot but now
for the stream density distribution. Here we see a very different result. The
stream density distribution depends quite strongly on softening, with smaller
softenings resulting in a shift of the entire stream density distribution
towards lower values. The effect is largest at small radii but is evident
throughout the main halo. In the inner halo the median stream density drops by
more than three orders of magnitude when the softening length is reduced by a
factor of 4.  The difference is even larger if we focus on the low-density
tail of the distribution and is still quite substantial in the high-density
tail. 

\begin{figure}
\center{
\includegraphics[width=0.435\textwidth]{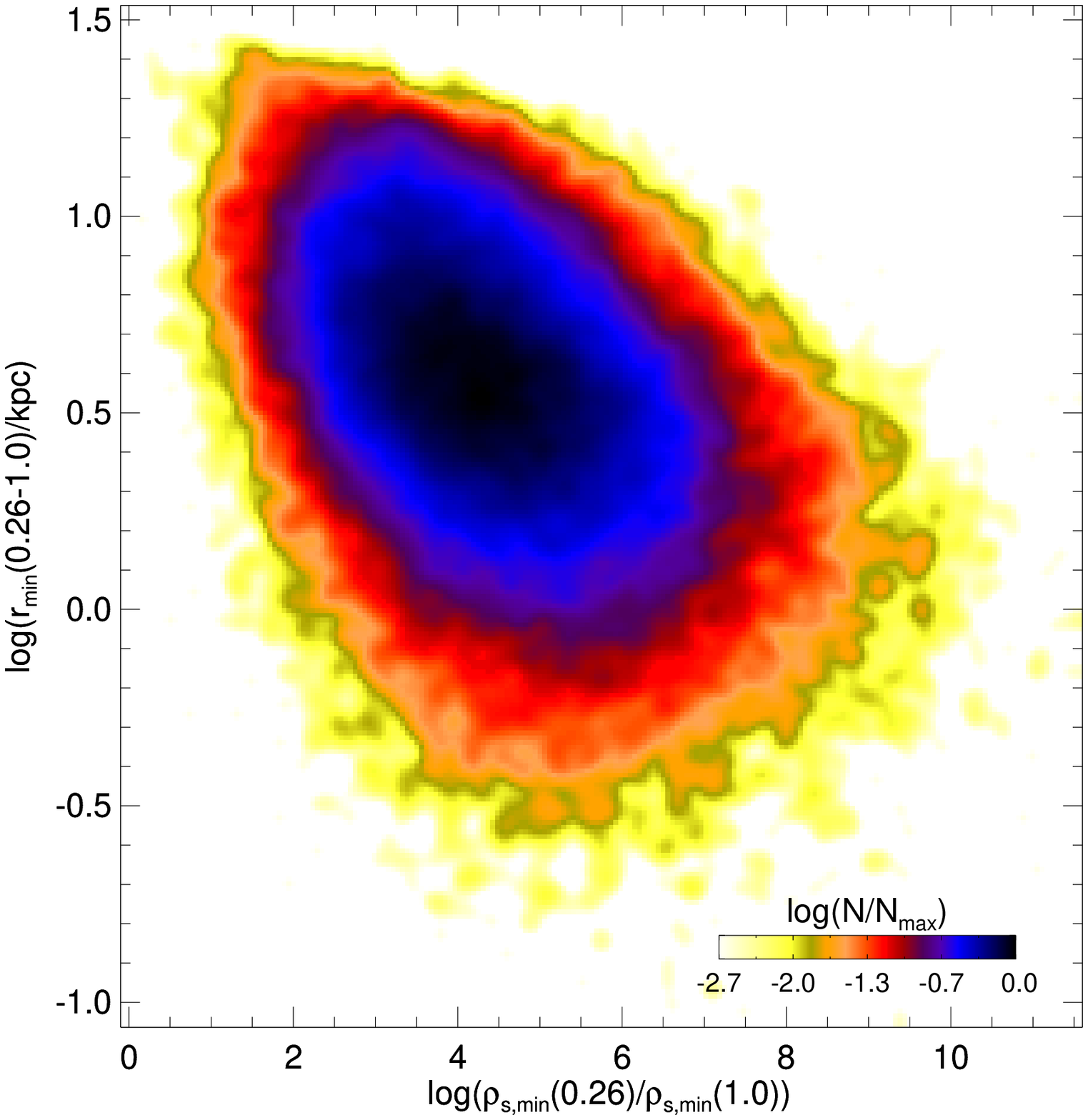}
\vskip 0.5cm
\includegraphics[width=0.44\textwidth]{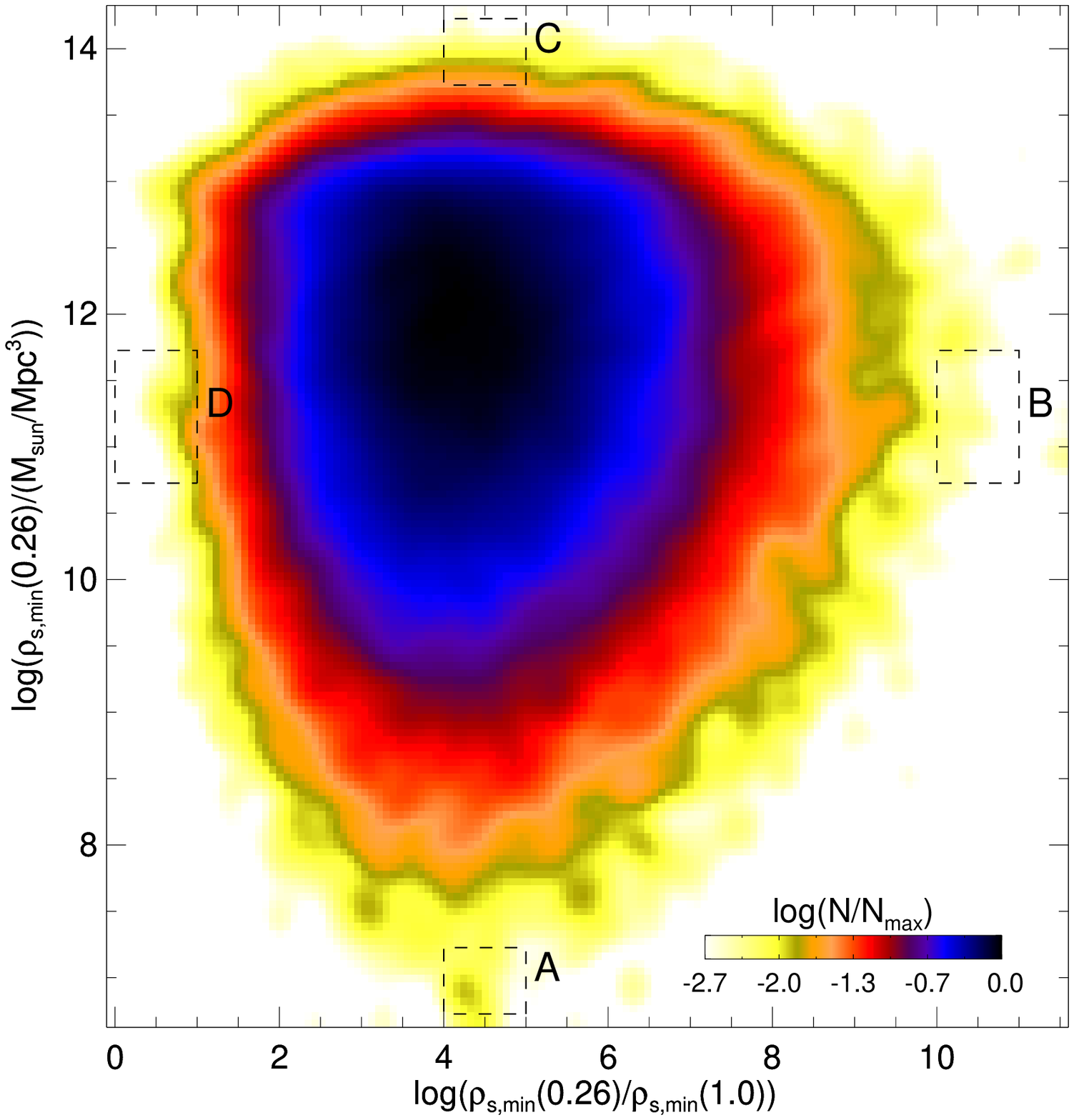}
}
\caption{Two-dimensional histograms of properties of a particle subset defined
  by the following criteria: halocentric distance at $z=0$ between
  $0.01~r_{200}$ and $r_{200}$; number of caustic passages from
  $a=0.26$ to $a=1$ between $38$ and $42$; already part of the main
  subhalo (i.e. not part of any identified subhalo) at $z=2.9$.  Top
    panel: Distribution in the $\rho_{\rm s,min}(0.26)/\rho_{\rm s,min}(1)$
  versus $r_{\rm min}(0.26-1)$ plane. Particles with smaller minimum
  radius over the $z=2.9$ to $z=0$ period typically decrease their stream
  density by a larger factor, i.e. $\rho_{\rm s,min}(0.26)/\rho_{\rm
    s,min}(1)$ tends to be larger for smaller $r_{\rm min}(0.26-1)$.
   Bottom panel: Distribution in the $\rho_{\rm s,min}(0.26)/\rho_{\rm
    s,min}(1)$ versus $\rho_{\rm s,min}(0.26)$ plane.  There is no correlation
  between the two quantities. Within this particle set, the change in stream
  density after $z=2.9$ is independent of a particle's prior history. Notice
  also that the distribution of the change in stream density is broader than
  the initial distribution, showing that the final stream density distribution
  is determined primarily by effects during the redshift interval considered.
  The small boxes labelled A to D select particles at the edges of the
  distribution. In Fig.~\ref{fig:time} we plot the evolution of stream density
  and halocentric distance for six particles chosen at random from each box.}
\label{fig:hist_count} 
\end{figure}

To understand the origin of this effect, we repeated the Aq-A-4 simulation
with half the fiducial softening and stored the particle data at all $\sim
7000$ time-steps between $z=4$ and $z=0$. This allows us to follow the orbit
and the phase-space density evolution of each particle in detail. We find the
$200$ most bound particles at $z=0$ and use their centre-of mass position to
define the halo centre at all earlier times. (We checked that they are all
indeed still very close to halo centre at $z=4$.) For every halo particle we
then construct trajectories of halocentric distance and fine-grained stream
density, and define a minimum physical halocentric distance over any chosen
scale factor interval ($r_{\rm min}(a_1-a_2)$), and a minimum physical stream
density achieved between $z=4$ and any later epoch ($\rho_{\rm
  s,min}(a)$). Fig.~\ref{fig:hist_peri} shows the $z=0$ distribution of all
particles in the $r/r_{200}$ versus $r_{\rm min}(0.2-1)$ plane. This
histogram shows the expected trend that particles at small $z=0$ radii tend to
have passed closer to halo centre in their past than particles at large $z=0$
radii. Interestingly, the typical minimum radius is more than an order of
magnitude smaller than the final radius, and the scatter in the distribution
is quite large. Thus at a third the virial radius (around 80~kpc) the
distribution peaks at a closest approach distance of about 3~kpc, thus within
our fiducial softening length. At a tenth the virial radius, most particles
have passed within 1.0~kpc of halo centre.  Thus, everywhere within the halo a
significant fraction of particles have passed through a region where tidal
forces are substantially affected by the softenings we are using. The tidal
field sources evolution in the GDE, so this affects our stream density
estimates. Specifically, particles feel stronger shearing at pericentre if the
softening length is smaller, so we expect larger stream density decrements for
smaller softenings.

To investigate this further, we select a subset of particles according to the
following criteria: halocentric distance at $z=0$ between $0.01~r_{200}$ and
$r_{200}$; caustic count increment from $a=0.26$ to $a=1$ between $38$ and
$42$; part of the main halo (i.e. not part of any resolved subhalo) at
$a=0.26$. This selects a set of particles with similar orbital periods which
have all been part of the main body of the system since $z=2.9$. The top panel
of Fig.~\ref{fig:hist_count} shows the distribution of these particles in the
$\rho_{\rm s,min}(0.26)/\rho_{\rm s,min}(1)$ versus $r_{\rm min}(0.26-1)$
plane.  The ratio $\rho_{\rm s,min}(0.26)/\rho_{\rm s,min}(1)$ indicates the
decrease in minimal physical stream density between $z=2.9$ and $z=0$, which,
as expected, is strongly correlated with minimum halocentric
distance. Particles with stream density changes substantially smaller than the
median have typically never been closer to the centre than 2 or 3~kpc, while
particles with substantially larger than average changes in stream density
have almost all been within 3~kpc of halo centre, many within 1~kpc.  The
bottom panel of Fig.~\ref{fig:hist_count} shows the same particle set in the
$\rho_{\rm s,min}(0.26)/\rho_{\rm s,min}(1)$ versus $\rho_{\rm s,min}(0.26)$
plane, i.e. the drop in physical stream density from $z=2.9$ to $z=0$ versus
the initial stream density at $z=2.9$. There is no correlation between these
quantities, showing that the change in stream density after $z=2.9$ is
independent of what happened to the particle at earlier times. In addition,
the distribution of stream density changes is broader than the initial
distribution, showing that the main features of the final distribution were
established over the time interval considered.

Together these plots show that, for this particle subset, low stream densities
at $z=0$ are associated almost exclusively with particles that pass within our
fiducial softening length of the centre some time after $z=2.9$. Even in the
main part of the stream density distribution about half of the particles pass
within 3~kpc of halo centre. As we will see below, the time-average
halocentric distance of these particles is about 40~kpc. This explains why
changing our softening between 1.7 and 6.8~kpc has effects out to large radii,
as seen in the stream density distributions of Fig.~\ref{fig:profile_check},
and why the effects are stronger near halo centre and in the low-density tail
of the distribution.

\begin{figure*}
\center{
\includegraphics[width=0.49\textwidth]{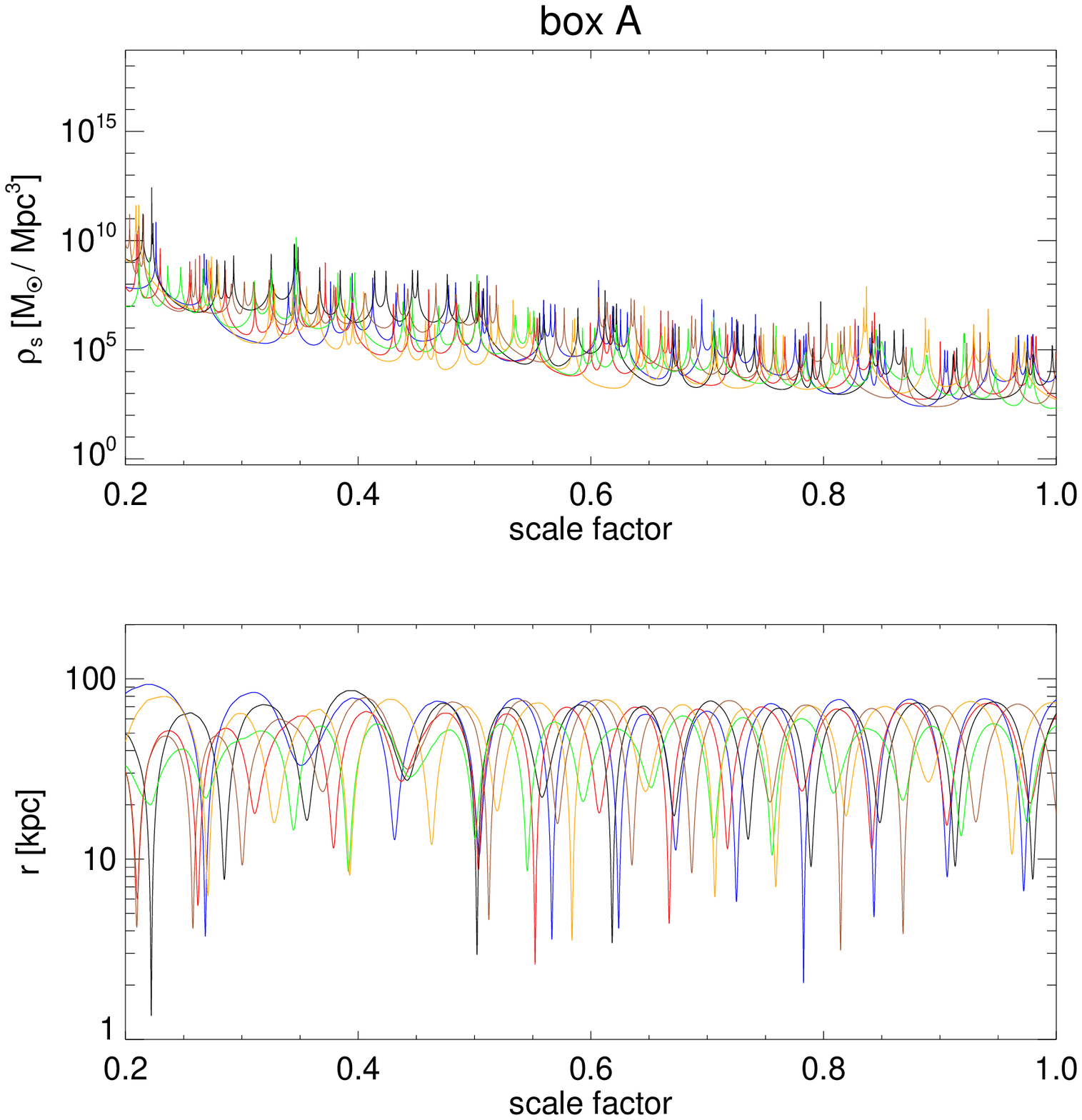}
\includegraphics[width=0.49\textwidth]{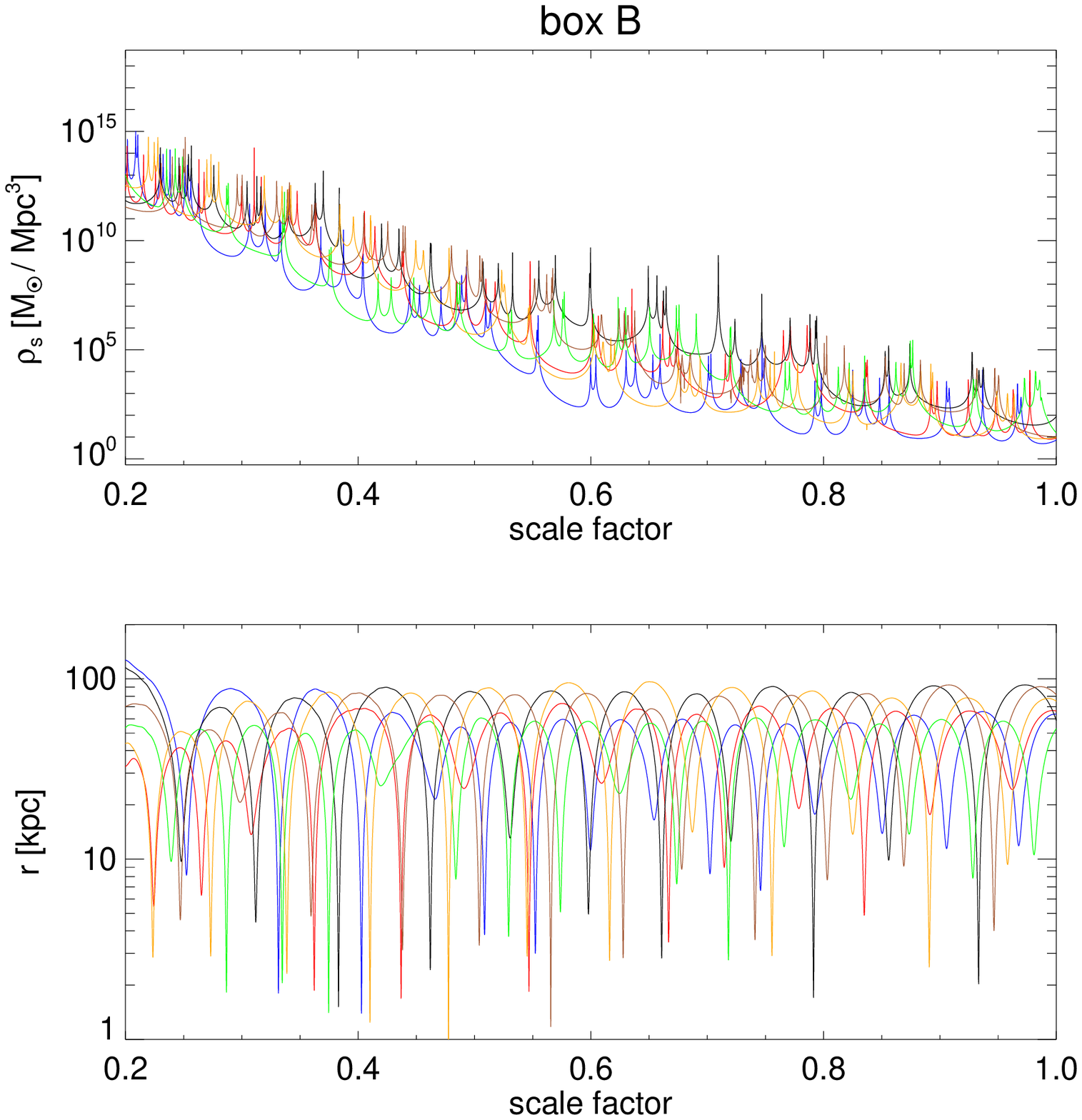}
\includegraphics[width=0.49\textwidth]{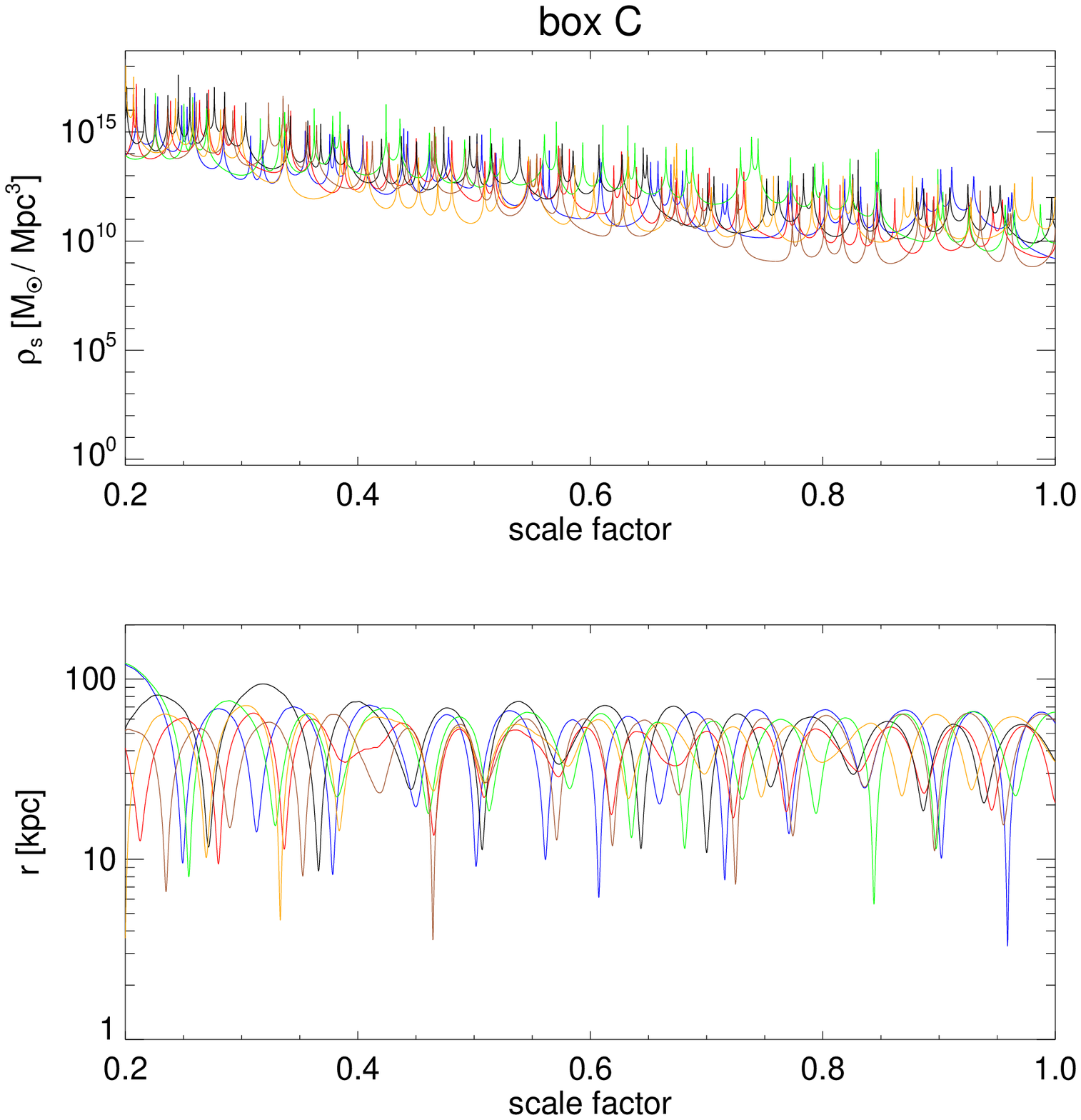}
\includegraphics[width=0.49\textwidth]{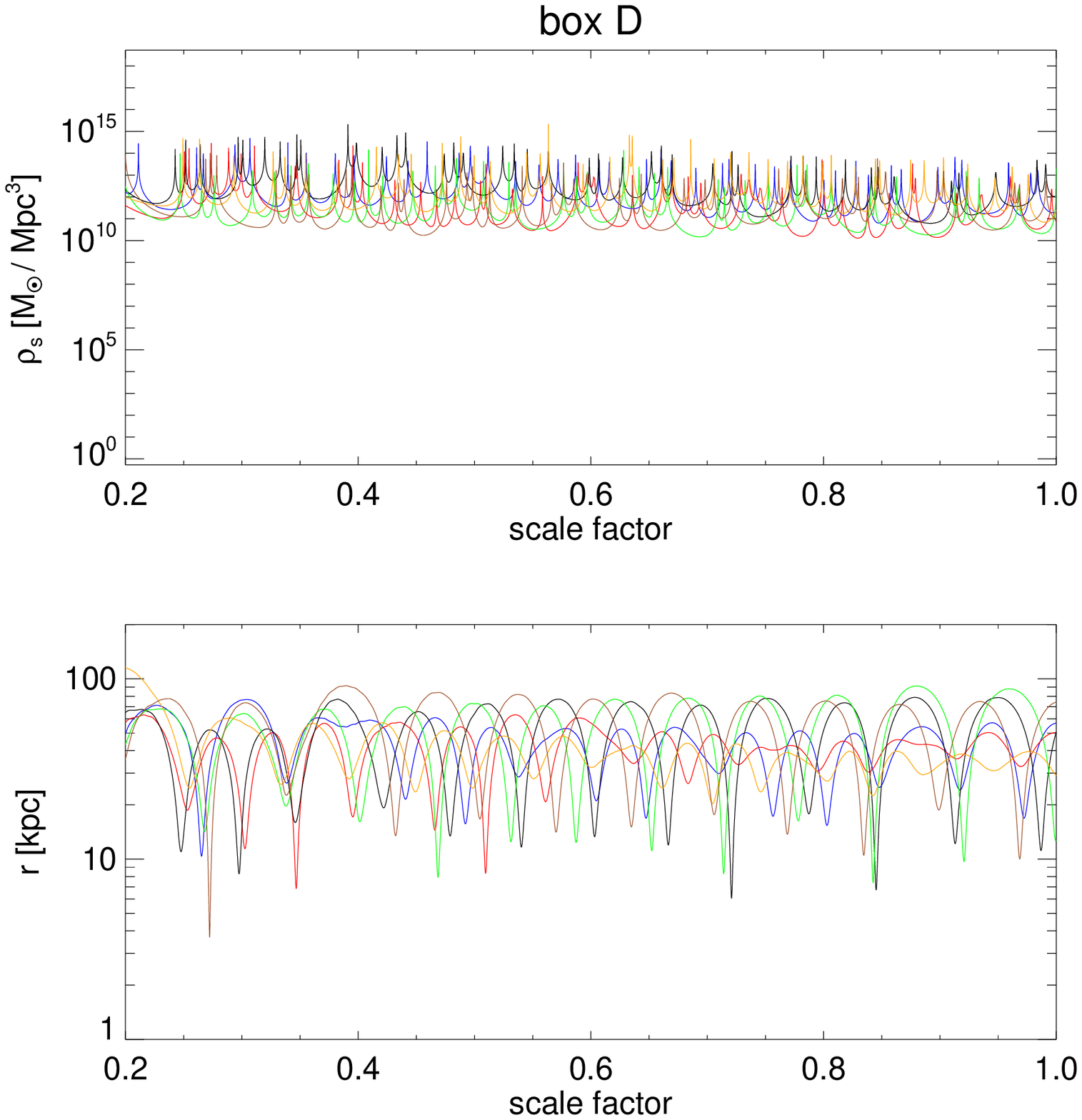}
}
\caption{ Stream density and halocentric distance (both physical) as a
  function of scale factor for six particles randomly selected from each of
  the regions labelled A to D in the bottom panel of
  Fig.~\ref{fig:hist_count}. Each particle has its own colour which
  corresponds in each pair of plots.} 
\label{fig:time} 
\end{figure*}

The stored data for our test simulation allow us to trace the detailed
evolution of stream density for individual particles. We illustrate this for a
few particles in the extreme regions of the distribution labelled A to D in
the bottom panel of Fig.~\ref{fig:hist_count}.  We pick $6$ particles at
random from each region and plot in Fig.~\ref{fig:time} their physical stream
density and physical halocentric distance as a function of scale factor. These
plots demonstrate that particles in fully general CDM halos show quite similar
stream density behaviour to those in simpler halo models: caustic ``spikes''
occur several times per orbit at turning points of the fundamental
oscillations, and secular evolution is evident in the steady decrease in the
minimum stream densities achieved between caustic passages. This resembles the
isolated halo results in \cite{2008MNRAS.385..236V}, showing that our GDE
scheme works correctly in the cosmological setup of this paper. The selection
criteria for our particle subset are evident in the similar apocentric
distances and similar caustic and pericentric passage numbers of all orbits,
despite their very different initial and final stream densities.

The stream density decrease is different in the four regions, reflecting their
location in the $\rho_{\rm s,min}(0.26)/\rho_{\rm s,min}(1)$ versus $\rho_{\rm
  s,min}(0.26)$ plane. Our GDE approach finds stream densities spanning almost
$18$ orders of magnitude for orbits of very similar size and period.  The
orbits in regions A and C look similar both in their stream density and in
their radial orbit behaviour. The only difference is an offset of about 6
orders of magnitude in stream density which is preserved throughout the
evolution. This emphasises that evolution from $z=2.9$ to $z=0$ is independent
of what happened before $z=2.9$ if one restricts oneself to main halo
particles with similar orbital periods. The particles in boxes B and D also
have similar orbital periods, but they now have similar initial stream
densities and very different stream density evolution. Those in box D barely
change their stream density over the period shown, while the stream densities
of those in box B drop by about 11 orders of magnitude. Here there are clear
qualitative differences in orbital structure.  The particles in box D avoid
the halo centre and in some cases are almost circular, and their caustics are
relatively uniformly spaced in time and show rather little variation.  In
contrast, the particles in box B make much larger radial excursions, often
passing close to the centre, and their caustic structure is more variable with
caustic passages often occurring in groups. This suggest that further work on
the structure of the stream density evolution curves might allow
classification of the orbits into tubes, boxes and other families.

\end{appendix}

\label{lastpage}

\end{document}